\RequirePackage[hyphens]{url}
\documentclass[aps,apl,reprint,superscriptaddress,showpacs,amsmath, amssymb, prl, floatfix, longbibliography]{revtex4-2}
\usepackage{graphicx}
\usepackage{color}
\usepackage{latexsym}
\usepackage{amssymb}
\usepackage{bm}
\usepackage{dcolumn}
\usepackage{epstopdf}
\usepackage{color}
\usepackage{appendix, hyperref}
\usepackage[normalem]{ulem}

\begin{document}



\title{Electron-beam Floating-zone Refined UCoGe}
\author{K. E. Avers}
\affiliation{Los Alamos National Laboratory, Los Alamos, NM 87545, USA}
\affiliation{Department of Physics and Astronomy, Northwestern University, Evanston, IL, USA}
\affiliation{Center for Applied Physics \& Superconducting Technologies, Northwestern University, Evanston, IL, USA}
\author{M. D. Nguyen}
\affiliation{Department of Physics and Astronomy, Northwestern University, Evanston, IL, USA}
\author{J. W. Scott}
\affiliation{Department of Physics and Astronomy, Northwestern University, Evanston, IL, USA}
\author{A. M. Zimmerman}
\affiliation{Department of Physics and Astronomy, Northwestern University, Evanston, IL, USA}
\author{S. M. Thomas}
\affiliation{Los Alamos National Laboratory, Los Alamos, NM 87545, USA}
\author{P. F. S. Rosa}
\affiliation{Los Alamos National Laboratory, Los Alamos, NM 87545, USA}
\author{E. D. Bauer}
\affiliation{Los Alamos National Laboratory, Los Alamos, NM 87545, USA}
\author{J. D. Thompson}
\affiliation{Los Alamos National Laboratory, Los Alamos, NM 87545, USA}
\author{W. P. Halperin}
\affiliation{Department of Physics and Astronomy, Northwestern University, Evanston, IL, USA}
\begin{abstract}
The interplay between unconventional superconductivity and quantum critical ferromagnetism in the U-Ge compounds represents an open problem in strongly correlated electron systems. 
Sample quality can have a strong influence on both of these ordered states in the compound UCoGe, as is true for most unconventional superconductors. We report results of a new approach at UCoGe crystal growth using a floating-zone method with potential for improvements of sample quality and size as compared with traditional means such as Czochralski growth. 
Single crystals of the ferromagnetic superconductor UCoGe were produced using an ultra-high vacuum electron-beam floating-zone refining technique. Annealed single crystals show well-defined signatures of bulk ferromagnetism and superconductivity at $T_c \sim$2.6 K and $T_s \sim$0.55 K, respectively, in the resistivity and heat capacity. Scanning electron microscopy of samples with different surface treatments shows evidence of an off-stoichiometric uranium rich phase of UCoGe collected in cracks and voids that might be limiting sample quality.
\end{abstract}

\maketitle

\section{Introduction}
The phenomenon of superconductivity, the abrupt disappearance of electrical resistance below a material specific temperature, was discovered over 100 years ago in elemental mercury at 4.16 K \cite{Onnes_Proceedings_1911, Delft_EuroPhysNews_2011}. Fifty-six years later, the BCS theory provided a microscopic understanding in which conduction electrons with opposite momentum form bound states called Cooper pairs mediated by lattice vibrations \cite{Bardeen_PhysicalReview_1957}.
It was thought that superconductivity was a solved problem until advances in sample growth techniques, such as flux growth, Czochralski, and solid state reaction techniques, lead to the discovery of high-temperature cuprate superconductors \cite{Bednorz_ZPB_1986, Wu_PRL_1987} and heavy fermion superconductors \cite{Steglich_PRL_1979} in which the microscopic origin of the pairing mechanism is not driven by phonons. 
In particular, uranium containing heavy fermion superconductors exhibit exotic behavior including time-reversal symmetry breaking in the superconducting state as reported for UPt$_3$ \cite{Schemm_Science_2014, Avers_NaturePhys_2020} and a hidden order state coexisting with time-reversal symmetry breaking superconductivity in URu$_2$Si$_2$ \cite{Wartenbe_PRB_2019, Schemm_PRB_2015}.
Another exciting feature is that they offer some of the clearest examples of systems with multiple superconducting phases, with both the aforementioned UPt$_3$ \cite{Adenwalla_PRL_1990} and Th doped UBe$_{13}$ \cite{Stewart_LowTempPhys_2019} demonstrating this unambiguously.
Although many types of superconducting compounds are known, almost all of them share a common trait: antagonism to external magnetic field \cite{Werthamer_PRL_1966}, magnetic impurities \cite{Suhl_PhysAndChemOfSolids_1959, Matthias_PRL_1958}, and ferromagnetic order \cite{Liu_PRB_2017}.
The curious exception to this rule at present is limited to the relatively new compounds UGe$_2$ \cite{Saxena_Nature_2000}, UCoGe \cite{Huy_PRL_2008}, URhGe \cite{Aoki_Nature_2001}, and possibly UTe$_2$ \cite{Ran_Science_2019, Aoki2_JPSJ_2019} in which ferromagnetic order, or rather ferromagnetic critical fluctuations \cite{Wu_NatComm_2017}, are responsible for superconductivity. We emphasize that the ferromagnetic fluctuation issue is an open question in UTe$_2$ \cite{Sundar_PRL_2019, Tokunaga_JPSJ_2019} due to recent neutron scattering experiments showing that antiferromagnetic ones are dominant \cite{Duan_Arxiv_2020}, which is also supported by recent hydrostatic pressure work \cite{Thomas_SciAdv_2020}. Despite all these materials having orthorhombic crystal structures \cite{Aoki_JPSJ_2019}, each have their own unique behaviors: UGe$_2$ exhibits crossover behavior between two ferromagnetic phases with a different ordered moment \cite{Pfleiderer_PRL_2002}, URhGe has a magnetic field induced superconducting phase disconnected from its zero field one \cite{Levy_Science_2005}, and UCoGe shows evidence of a first order ferromagnetic phase transition \cite{Ohta_JPSJ_2010}.

\begin{figure}[h!]
	\centering
	\hspace{1em}
	\includegraphics[width=0.95\columnwidth]{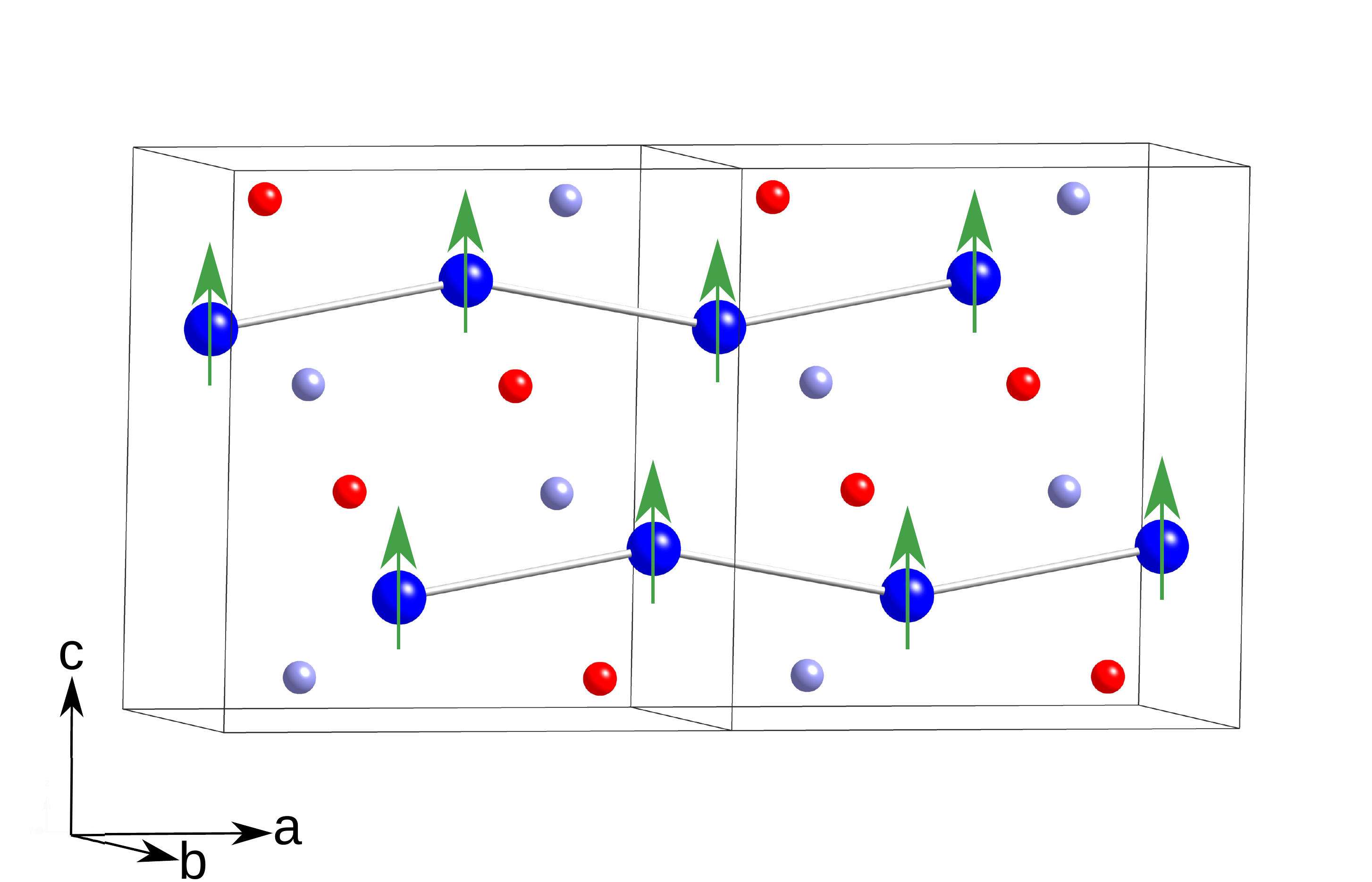}
	\caption{Crystal structure of UCoGe.  Uranium atoms: large blue balls; cobalt atoms: light blue balls; germanium atoms: red balls. The lines between U atoms are drawn to emphasize the zig-zag arrangement. The green arrows show the Ising ferromagnetic moments on the uranium sites.}
	\label{fig:Crystal}
\end{figure}

The U-Co-Ge ternary system includes a number of compounds such as the room-temperature ferromagnet U$_2$Co$_{17-y}$Ge$_y$ \cite{Chevalier_AlloysAndCompounds_1996}, the Pauli paramagnet U$_3$Co$_{12-x}$Ge$_4$ \cite{Soude_SolidStateChem_2010}, and the heavy fermion low temperature ferromagnet U$_3$Co$_2$Ge$_7$ \cite{Bobev_SolidStateChem_2007}. Its metallurgy is largely uninvestigated, which is a potential hindrance to the improvement of UCoGe single crystals. 
UCoGe may have a small peritectic window above 1300 $^\circ$C and up to its melting point at $\sim$ 1360 $^\circ$C \cite{Eric_Private_2018}, making it feasible to obtain large single crystals from a high temperature stoichiometric melt. 
The majority of reported single crystal samples have been grown by Czochralski \cite{Huy_PRL_2008, Wu_PRB_2018, Hattori_PRL_2012}, with the exception of one growth attempt with optical floating-zone refining \cite{Pospisil_JPSJ_2014}. 
This motivated our attempt to grow UCoGe with an ultra-high vacuum (UHV), electron-beam floating-zone technique.  
In previous work with this method, some of the highest quality bulk crystals of UPt$_3$ were reported \cite{KyciaThesis, Kycia_PRB_1998, Joynt_RevModPhys_2002}. However, maintenance of stoichiometry in vacuum with volatile components, such as Ge and Co, was a potential drawback. We report here that this drawback is not insurmountable.
UCoGe ideally crystallizes in the orthorhombic TiNiSi structure as shown in Fig.\,\ref{fig:Crystal} with the ferromagnetic Ising magnetic moments pointing along the c-axis below $T_c$= 2.6 K \cite{Aoki_JPSJ_2019}.
UCoGe may also crystallize in the related disordered CeCu$_2$ structure which is characterized by random site mixing between the Co and Ge sites \cite{Gasparini_LowTempPhys_2010}.

In this work, we detail the growth and characterization of UHV electron-beam floating-zone refined UCoGe single crystals. The crystals were examined under a variety of conditions (unannealed, annealed, acid-etched, polished, etc) using scanning electron microscopy (SEM), low-temperature electrical resistivity, heat capacity and magnetization. We find our crystals are of high quality and display evidence of bulk superconductivity and ferromagnetism, which we will show are consistent with previously reported high quality UCoGe single crystals. By comparing results of these measurements under different sample conditions we have found evidence of a yet undiscovered/unreported off-stoichiometric uranium rich phase of UCoGe, referred to as the low temperature phase (LTP), that coexists within cracks and voids inside the UCoGe single crystal. This LTP may be a key reason for the difficulty in improving the quality of UCoGe single crystals. Our work may guide future efforts in improving the quality of UCoGe single crystals, and hence enabling a better understanding of the exotic phenomena of ferromagnetic critical fluctuation mediated superconductivity.

\section{Methods}
Near-stoichiometric amounts of U (electromigration purified to 99.99 \% purity), Co (99.99 \%), and Ge (99.9999\%) elements were arc-melted into 4 individual buttons of UCoGe on a water cooled copper hearth under $\sim$20 torr of Ar (99.999 \%). The average stoichiometry of the buttons determined by the initial mass of the constituent elements is 33.48 \% U, 33.16 \% Co, and 33.36 \% Ge. All percentages in this work are atomic percent (at. \%), not mass percent.
Before each arc-melting operation, a getter made of uranium was heated into the liquid state in order to absorb oxygen. 
The buttons were arc-melted into two final ingots in a long narrow trough formed in the hearth. The ingots had diameters of $\sim$ 6 mm and lengths of $\sim$ 150 mm and $\sim$ 65 mm with a combined mass of $\sim$ 57 g. 
The ingots were quite brittle and tended to crack and eventually break if mishandled, requiring a few attempts to optimize their casting.
During arc-melting it was observed that the ingots would not remain straight, but rather would bend upward away from the hearth suggesting there exists a high temperature structural transition in the solid phase as it cools. 
Effects of this high-temperature phase have been reported, with prior samples becoming distorted when annealed close to the melting temperature of UCoGe \cite{Taupin_PRB_2014}.
It was also observed that small flakes of a high temperature solid phase, suspected to be a uranium oxide phase based on later SEM observations, floated on top of the molten parts of the ingot both during arc-melting and zone refining. 

\begin{figure}[h!]
	\centering
	\hspace{1em}
	\includegraphics[width=0.95\columnwidth]{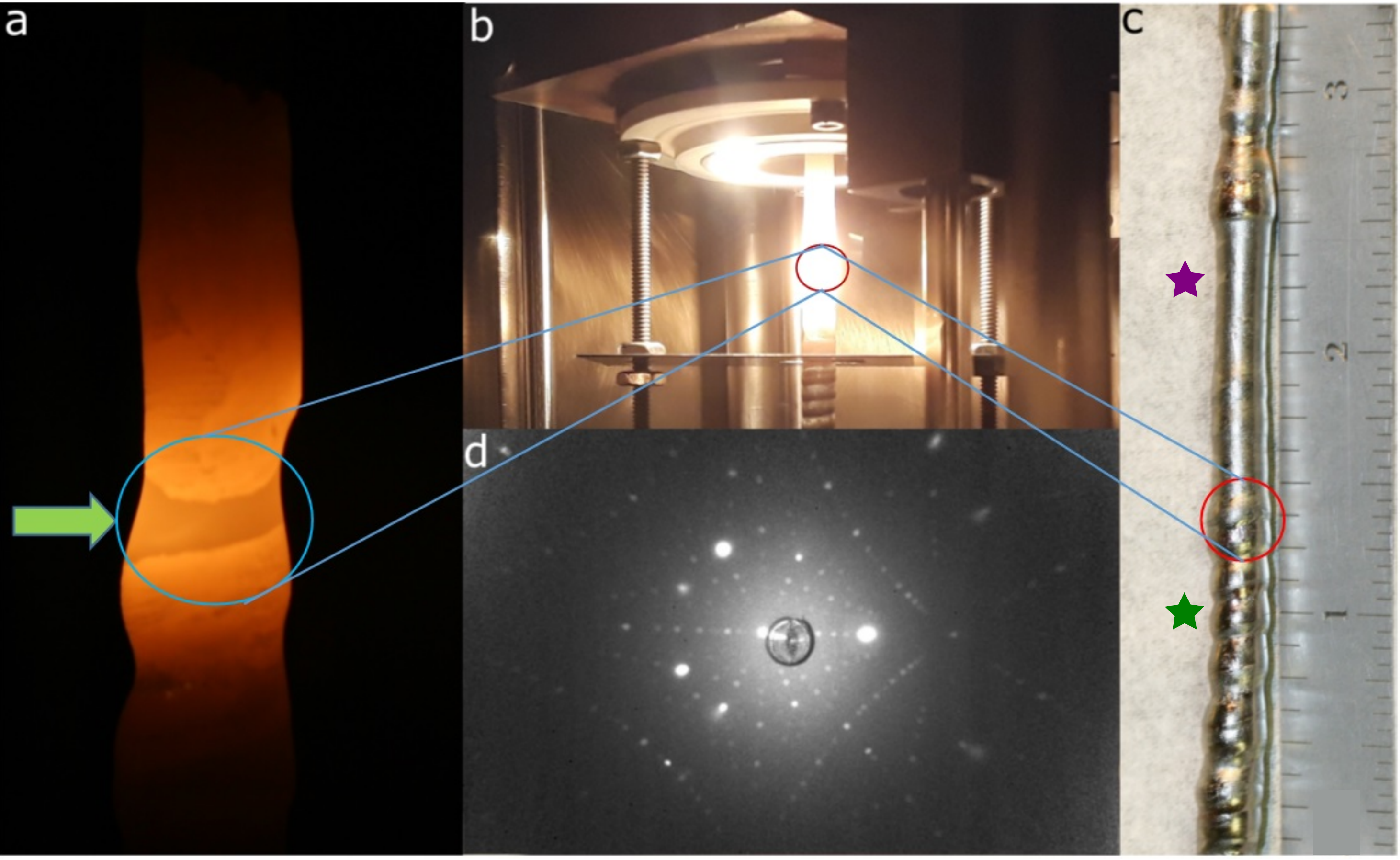}
	\caption{(a): Photograph of the molten floating-zone during growth of UCoGe.  The green arrow indicates the fully molten floating-zone that was being swept upward. (b): View of the ingot being zone refined showing the circular electron gun filament  housed in a water cooled copper translation stage, and the molybdenum grounding plate used to focus the electron-beam held below by two threaded shafts. (c): Picture of UCoGe sample after three zone refining passes over a $\approx 75$\,mm (3 inches on the ruler) long segment of the sample with the approximate location of S1 and S2 shown by the green and purple stars, respectively. (d): Laue x-ray diffraction image of an aligned piece of the UCoGe zone refined sample. The diffraction pattern shows the sample is aligned close to the [010] (b-axis) direction demonstrating the single crystalline nature of the zone refined sample. 
	}
	\label{fig:Ingot}
\end{figure}

The final ingots were mounted vertically in a UHV electron-beam floating-zone refining apparatus, clamped at their ends between molybdenum fixtures and set screws, with the longer secured ingot above and the shorter ingot below. The specifics of the zone refining apparatus has been previously described in Ref. \citenum{KyciaThesis}. 
The vacuum chamber was evacuated to $\sim$8 nTorr via a high temperature bake out procedure using a diffusion pump with a liquid nitrogen baffle to prevent oil vapor contamination.
The ingots were then joined in-situ using the electron gun, and a $\sim$75 mm length of the ingot was refined 3 times by translating the electron gun upwards with a rate of $\sim$15 mm/hr as shown in Fig.\,\ref{fig:Ingot} (a) and (b). 
		
		During zone refining a liquid nitrogen Meissner trap was used to capture volatile impurities. 
In addition to any undesired volatile impurities, the viewing ports became darkened due to the relatively high vapor pressure of the Co and Ge compared to U.
It was later shown via energy-dispersive x-ray spectroscopy (EDS) that the stoichiometry may have changed. The resulting ingot, after being removed from the chamber, is shown in Fig.\,\ref{fig:Ingot} (c).
The variation of diameter in the lower parts of the ingot is due to adjustments of electron-beam power as well as vertical adjustments of the lower ingot necessary to maintain a stable floating-zone. 
In contrast, the upper part of the ingot required fewer adjustments resulting in a more uniform cross section and smoother as-grown surface.
The discolorations are impurities that floated to the surface with the greater amount of discoloration near the top indicating that the zone refining process was able to segregate impurities to the top of the ingot.
The impurities and surface structure hindered Laue x-ray diffraction on the as-cast surface resulting in patterns with smeared Bragg peaks, or even no observable Bragg peaks at all. This necessitated spark cutting, polishing, and orienting for the Laue pattern in Fig.\,\ref{fig:Ingot} (d) aligned along the [010] (b-axis) direction.
Subsequent Laue surveys suggested that the ingot did not become a large single crystal, but rather multiple macroscopic crystal grains.
This was determined by taking multiple Laue exposures at different locations along the length of the ingot where we found a uniform Laue pattern, and hence a crystal grain, that persisted over at least $\sim$ 10 mm.
Once an appropriate region belonging to a single crystal was found, an oriented wafer was spark cut from that region of the ingot, polished, and further Laue surveys conducted to determine if the interior region was single crystalline.
It was observed that neither the a, b, nor c-axes coincided with the growth direction during zone refining which required cuts to be made at reasonably large inclination angles ($\sim$30$^\circ$) relative to the cylinder axis of the ingot.

Once the wafer was determined to be a suitable single crystal grain, it was further oriented and cut into needle shaped samples for further characterization.
Two needles, labeled S1 and S2, were oriented and spark cut to 3.10 mm $\times$ 0.96 mm $\times$ 1.10 mm, and 4.30 mm $\times$ 1.10 mm $\times$ 1.30 mm, respectively, along the a, b, and c-axes. 
S1 was taken from the lower part of the ingot while S2 was taken closer to the top, with the approximate location of both needles indicated in Fig.\,\ref{fig:Ingot} (c). Note that substantial diameter variation in the region of S1, while the diameter of S2 is relatively smooth.
The b-axis normal surface  of these needles was polished to a mirror shine using 1200 grit SiC sandpaper for the x-ray Laue diffraction measurements performed on the parent wafers.

In the following text, we will discuss the samples in three preparation conditions, hereafter denoted X, Y, and Z.
The initial sample conditions after zone refinings will be referred to as X or ``unannealed''. Samples annealed under UHV conditions for 2 weeks at 900 $^\circ$C while resting on polycrystalline UCoGe wafers, but with no further treatment, will be referred to as Y or simply ``annealed''.
After annealing, it was observed that a liquified phase collected on the polished (ac-plane) surfaces, but not the spark cut (ab-plane) surfaces. 
This phase will be referred to as the `low temperature phase' (LTP), due to its relatively low melting point. Powder X-ray diffraction (not shown) did not indicate existence of any phases other than UCoGe in either arc-melted or zone refined material, so the identity of this phase remains unknown, although it is clearly visible optically and in the SEM data. Due to the LTP, the needles were further polished to a final thickness along the b-axis direction of 0.40 mm and 0.19 mm for S1 and S2, respectively.
This final sample condition will be referred to as Z or ``annealed/polished''.
In addition, when discussing results pertaining to the outer diameter of the ingot it will be referred to as the ``as-grown'' surface in contrast to discussing surfaces from the interior that were spark cut and polished. 


SEM images were collected using a Hitachi S-3400N-II with an attached Oxford INCAx-act SDD EDS to determine relative elemental composition via the AZtec software package.
All elemental percentages have an error of $\sim$1 at.\% as reported by the software, although we comment that since \emph{in situ} elemental references were not used the actual error may be larger.
Electrical resistivity measurements were performed using a standard four-point method down to 425~mK for unannealed (X) S1 and for annealed (Y) S1 and S2.
The temperature was measured by a germanium resistance thermometer that was calibrated below 45 K.
Heat capacity measurements were performed using a Quantum Design PPMS.
After the samples were polished along the b-axis direction to their final thickness as annealed/polished (Z) the resistivity was remeasured without the surface influence from the LTP.
This second series of resistivity measurements were made with a Quantum Design PPMS.
For the resistivity data that will be presented the excitation current was 0.6 mA for unannealed (X) S1 X, 0.1 mA for annealed (Y) S1, 0.5 mA for annealed (Y) S2, and 0.1 mA for annealed/polished (Z) S1 and S2. Magnetization measurements were done using on a Quantum Design MPMS.
\section{Results}

\begin{figure}[h!]
	\centering
	\hspace{1em}
	\includegraphics[width=0.95\columnwidth]{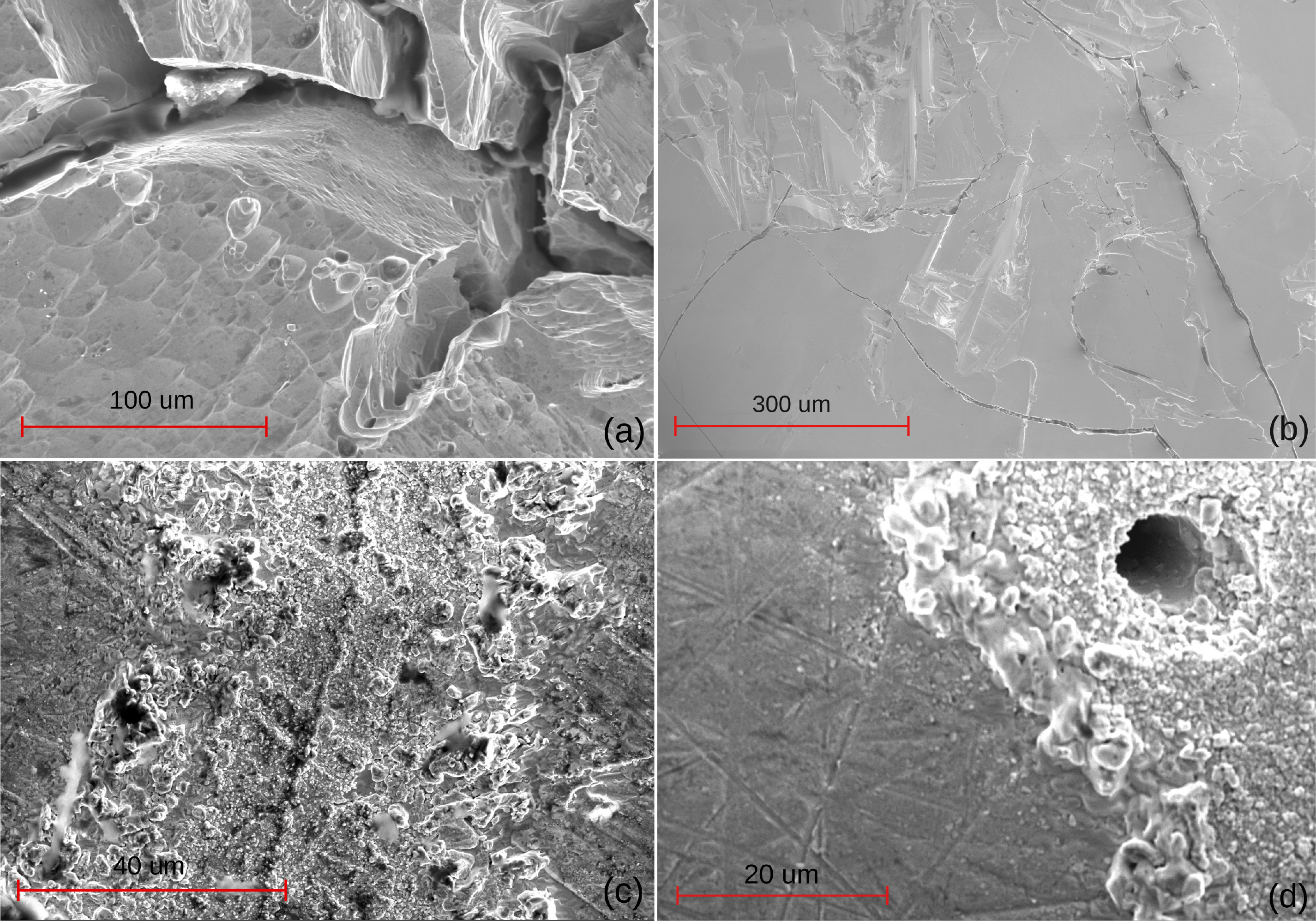}
	\caption{SEM images taken from various zone refined UCoGe samples near the bottom of the ingot at different stages of treatment. (a) SEM image of warm HCl+nitric acid etch displaying a large crack. (b) SEM image of the as-grown surface displaying micro-structure of the unknown high temperature phase observed during the growth process, suspected to be a uranium oxide due to the oxygen detected in EDS on this surface. (c), (d) SEM image of a surface that was polished to a mirror finish, and then annealed at 900 $^\circ$C for 2 weeks (condition Y). The LTP liquified and collected on the polished surface via either large cracks (c), or from small holes (d).}
	\label{fig:SEM}
\end{figure}

SEM images from UCoGe under various treatment  conditions are presented in Fig.\,\ref{fig:SEM}. These images were collected from wafers near S1 (Figs. \ref{fig:SEM} (a), (b)) or in the case of Fig. \ref{fig:SEM} (c) and Fig. \ref{fig:SEM} (d) directly from S1. In general, SEM was not performed on pieces oriented on a crystallographic axis, but rather circular wafers cut from the main ingot with the cut surface normal along the growth direction. 
Fig.\,\ref{fig:SEM} (a) shows an interior surface that was polished, and then etched in a warm mixture of nitric and hydrochloric acid where the etching produced a large crack in the surface. 
This suggests that a metallurgical phase other than UCoGe is present that is more susceptible to the etching and is likely to be the aforementioned LTP, even though such a phase could not be detected in powder x-ray diffraction.
We acknowledge that this crack could instead have started as a dislocation or grain boundary. This would also provide an energetically favorable location for the chemical reaction, but without detailed transmission electron microscopy results it is difficult to comment on the presence of larger scale structural defects in UCoGe. 
The interior region had EDS spectra consistent with a U:Co:Ge molar ratio of $\sim$ 1:1:1 ($\pm$0.03, or 1\%). 
The as-grown surface is shown in Fig.\,\ref{fig:SEM} (b) showing a micro-structure that formed at the surface during growth. 
Unlike SEM taken from the etched surface of interior regions, the as-grown surface showed regions with an oxygen signal in the EDS spectra, suggesting that a uranium oxide segregated to the outside diameter of the ingot during growth.
This is consistent with the aforementioned solid flakes we observed on the molten surface during arc-melting and zone refining.
This slight oxide contamination is likely due to natural oxidation when the stock material was exposed to air before and after arc-melting attempts, as well as the $\sim$ 3 months the ingots were left exposed to air between the final arc-melting and zone refining. 
The as-grown surface also had measurable regions of $\sim$ 1:1:1 $\pm$ 0.03 free of oxygen, indicating that the oxidation is not completely coating the surface.
The surface structure and oxide made Laue difficult when taken directly from the as-grown surface. 
Both arc-melted and zone refined polished surfaces (not shown) are completely featureless other than scratch marks from polishing, and also had EDS spectra consistent with UCoGe. Phase boundaries between the UCoGe and any secondary phases could not be resolved.
In contrast Fig.\,\ref{fig:SEM} (c) and (d) show the aftermath of annealing on polished surfaces. The LTP liquified and flowed through micro-cracks (c), and holes in the material (d). Upon solidification the LTP formed small particles, with a larger concentration of particles near the cracks and holes.

The average elemental composition for each of the different sample treatments is reported in table \ref{tab:EDS}.
The arc-melted results are very similar to the percentages calculated from the initial mass of the constituent elements used, suggesting that there is minimal evaporation during arc-melting, and that the internal calibration used by the AZtec software is reasonably accurate for the UCoGe system.
After zone refining, the amount of uranium detected increased to 35 \% consistent with the minor, but noticeable, evaporation of Co and Ge observed during zone refining. After acid etching, the surface uranium rich regions were removed and compositions of U, and Co were selectively reduced such that the surface stoichiometry was close to ideal. Nonetheless, there were some spectra collected that had \% Ge as high as $\sim$43 \% consistent with Ge being chemically inert compared to U and Co. In contrast, the after zone refined and arc-melting SEM observations had minimal variance from site to site of a few percent, within error.
Most of the measurements are within a reasonable error consistent with on-stoichiometric UCoGe, but the unannealed as-grown surface condition X and zone refined annealed condition Y surface are uranium rich. 
The as-grown surface became rich in uranium due to the segregation of uranium oxide(s) to the outer diameter of the ingot during growth, although it is also possible that the as-grown surface was depleted of Co and Ge since that is where the evaporation occurs. 
It is harder to confirm the latter as we also observed relatively oxygen free regions that had percentages consistent with UCoGe. 
However, the annealed surface also had a noticeable increase in U, with no region that could be identified with percentages consistent with ideal stoichiometry, unlike all the other EDS measurements.
The highest recorded uranium concentration was $\sim$ 70 \% directly from one of the large particles of LTP that formed. This indicates that this LTP is richer in U compared to UCoGe. In contrast to our observations, there have been previous secondary phases resolved in SEM on UCoGe such as elemental U, U$_3$O$_8$ and U$_3$Co$_4$Ge$_7$ \cite{Pospisil_JPSJ_2014}. However, all three of these previous phases are unlikely to be the LTP, as both U and U$_3$O$_8$ melt at higher temperatures than the 900 $^\circ$C annealing temperature, whereas U$_3$Co$_4$Ge$_7$ has the incorrect stoichiometry to be consistent with the increase of U concentration observed in the condition Y results in table \ref{tab:EDS}.

\begin{table}[h!]
  \begin{center}
    \begin{tabular}{l|c|c|c} 
		\hline
		\hline
      Condition & \% U & \% Co & \% Ge \\
      \hline
      arc-melted & 34$\pm$1 & 33$\pm$1 & 33$\pm$1 \\
      zone refined polished (X) & 35$\pm$1 & 32$\pm$1 & 33$\pm$1 \\
      zone refined acid etched (X) & 33$\pm$1 & 32$\pm$1 & 35$\pm$1 \\
			zone refined as-grown surface (X) & 63$\pm$1 & 17$\pm$1 & 20$\pm$1 \\
			zone refined annealed (Y) & 52$\pm$1 & 26$\pm$1 & 22$\pm$1 \\
					\hline
		\hline
    \end{tabular}
  \end{center}
      \caption{Average atomic percentages (at. \%) of the major elements at various stages of treatment. With the exception of the arc-melted row all entries are from material near S1. All entries were from interior regions of the crystal that were spark cut and polished except the zone refined as-grown surface entry. Most surfaces had ratios consistent with UCoGe, except the zone refined as-grown surface due to uranium oxide segregating to the outer surface of the ingot during growth, and the zone refined annealed surface due to the liquefaction of the LTP that collected on the polished surface. The 1 \% error is reported by the AZtec software, but this should be interpreted as a lower bound given the lack of reference sources used for calibration.}
					\label{tab:EDS}
\end{table} 

\begin{figure}[h!]
	\centering
	\hspace{1em}
	\includegraphics[width=0.95\columnwidth]{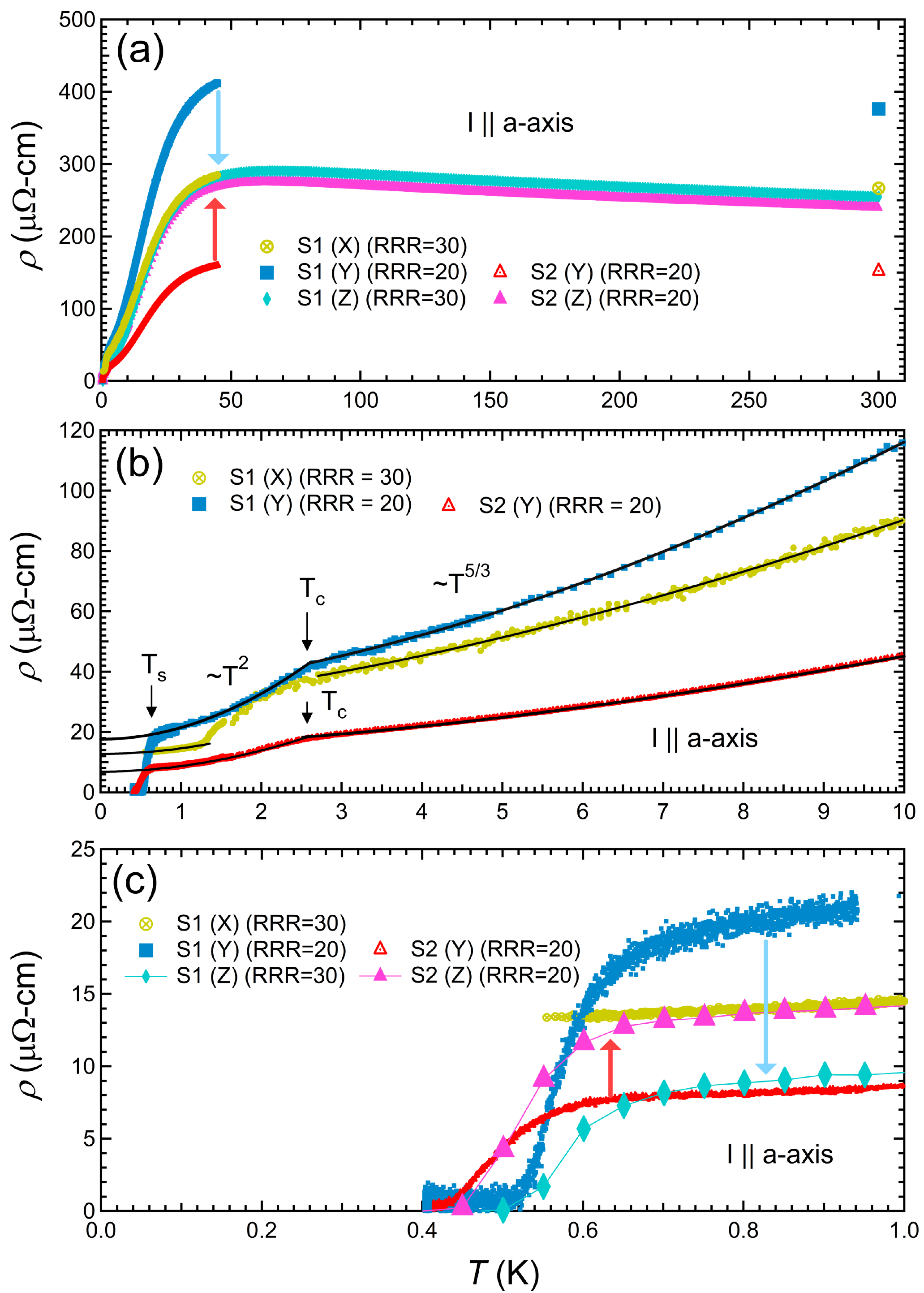}
	\caption{Electrical resistivity $\rho$ vs. temperature $T$ of zone refined UCoGe for current parallel to the a-axis (I$\parallel$a-axis) for the two samples S1 while unannealed (X), annealed (Y), and annealed/polished (Z) and S2 while annealed (Y), and annealed/polished (Z) from the bottom and top of ingot, respectively, from 0 K to 300 K (a), below 10 K (b), and below 1 K (b). Only annealed (Y) $\rho$ show deviations that can be ascribed to the LTP.
	(a) shows the difference that polishing the LTP off the electrical contact surface made in achieving consistent behavior between S1 and S2. The blue arrow shows that $\rho$ for S1 decreased at all temperatures, while the red arrow points out the opposite behavior for S2. Once polished (Z) both samples match quantitatively and qualitatively with the unannealed (X) S1 indicating that the damage from the the LTP is limited to the surface of the sample.
	(b) shows the ferromagnetic ($T_c$) and superconducting transitions ($T_s$) for the annealed (Y) S1 and S2, and the lack of superconductivity and broadened ferromagnetic transition in unannealed (X) S1. All solid curves are power law fits to $\rho=\rho_{0}+AT^\frac{5}{3}$ above $T_c$ or to $\rho=\rho_{0} +AT^2$ in the ferromagnetic state.
	(c) shows that the superconducting transition temperature for both samples remains the same between annealed (Y) and annealed/polished (Z), except afterward the sample with the higher zero resistance superconducting transition temperature of 0.52 K (S1) now has a lower $\rho(T)$ than the sample with the lower superconducting transition temperature of 0.44 K (S2).
	}
	\label{fig:rho}
\end{figure}

The temperature dependence of the resistivity, $\rho(T)$, taken from the ac-plane surface is shown in Fig.\,\ref{fig:rho} with electrical current applied parallel to the a-axis up to 300~K (a), 10~K (b), and 1~K (c).
As observed in Fig. \ref{fig:rho}(a), all data exhibit the feature of a local maximum in resistivity at roughly 65~K, known to be characteristic of the Kondo coherence effect \cite{hewson_1993}.
It is apparent that unannealed (X) and annealed/polished (Z) resistivities in Fig.\,\ref{fig:rho} (a) match each other quite well, except at the lowest temperatures in Fig.\,\ref{fig:rho} (c). The annealed (Y) behavior is an anomaly worth examining as the discrepancy is greater than that which can be explained by measurement error of the dimensions of the samples. 
The annealed (Y) S1 resistivity (blue squares) is substantially greater than unannealed (X) S1 resistivity (yellow circles) in Fig.\,\ref{fig:rho} (a) suggesting that the cracks and inhomogeneous distribution of the LTP shown in Fig.\,\ref{fig:SEM} (c) and (d) affects electrical transport.
It is evident that the annealed (Y) S1 resistivity is nearly a factor of 3 larger in magnitude compared to the S2 data (red triangles). After the samples were polished (Z) the S1 (turquoise diamonds) and S2 (purple triangles) magnitudes of $\rho(T)$ decreased and increased, respectively, as represented by the light blue arrow and red arrow.
In addition, the annealed/polished (Z) $\rho(T)$ for both S1 and S2 match the unannealed (X) S1 data at high temperatures demonstrating that most of the damage from the LTP is confined to the surface and is able to be removed. 

This limit of the influence of the LTP to the surface is supported in Fig. \ref{fig:rho} (b) by annealed (Y) S1 having a smaller RRR\,=\,$\rho(300 K)/\rho(T\rightarrow 0 K)$ of 20 compared to the unannealed (X) case with a RRR of 30, where the extrapolation is from the non-superconducting state.
The influence of the cracks and LTP on annealed (Y) S2 results in a much smaller $\rho(T\rightarrow 0 K)$, despite having the same RRR.
This suggests that the LTP is less resistive than the bulk UCoGe, and that electrical transport can be partially shorted depending on the exact location of the electrical contacts relative to the puddles of LTP.
The $\rho(T\rightarrow 0 K)$ for both annealed (Y) S1 and S2 are $\sim$ 18 $\mu \Omega$-cm and $\sim$ 7 $\mu \Omega$-cm respectively, while for S1 when unannealed (X), it is $\sim$ 13 $\mu \Omega$-cm.
The onset of superconductivity ($T_s$) for S1 is at $T_s \sim$ 0.61 K determined by linearly interpolating resistivity between the normal state and superconducting transition regions. 

As shown in Fig. \ref{fig:rho} (b), all three data sets have a $\sim T^\frac{5}{3}$ power law behavior above a Curie temperature ($T_c$) of $T_c \sim$2.6~K. 
Only when annealed (Y) do both samples show a well-defined power law of $\sim T^2$ in the ferromagnetic state, while unannealed (X) S1 has a very broadened ferromagnetic transition with no indication of superconductivity.
Both power law behaviors are consistent with a clean metallic ferromagnetic system dominated by electron-magnon scattering \cite{Brando_RevModPhys_2002}, and has been observed previously in other high quality UCoGe samples \cite{Huy_MagMaterials_2009}.
We note that $\sim T^2$ is also consistent with electron-electron scattering in a Fermi liquid, as has been observed in a large number of non-magnetic heavy fermion compounds \cite{Jacko_Nature_2009}. It is likely that both electron-magnon and electron-electron scattering are present in UCoGe, but that the former is dominant. That is indicated from the Kadowaki-Woods ratio, a comparison between the $\sim T^2$ resistivity coefficient ($\sim$ 2.5 $\mu \Omega$-cm/K$^2$ for our annealed/polished (Z) samples) and the heat capacity squared ($\sim$ (56)$^2$ mJ$^2$/mol$^2$-K$^4$ as estimated from heat capacity later), for UCoGe is $\sim$ 800 $\mu \Omega$-cm mol$^2$ K$^2$/J$^{2}$. 
This ratio is much larger than in heavy fermion superconductors ($\sim$ 10 $\mu \Omega$-cm mol$^2$ K$^2$/J$^{2}$) \cite{Jacko_Nature_2009} placing the two systems in distinctly different categories where the reasons for the enhanced resistivity coefficient and heat capacity are different. For further insight into this relation between resistivity and heat capacity in the ferromagnetic superconductors see Refs. \cite{Aoki_JPSJ_2014, Hardy_PRB_2011}. 

As can be seen in Fig. \ref{fig:rho} (c), the annealed (Y) samples have zero resistance superconducting temperatures of $T_s$~$\sim$ 0.52 K, and $\sim$ 0.44 K for S1 and S2, respectively. 
There is no resolvable change in the superconducting transition temperatures after polishing (Z); both S1 and S2 maintain the same zero resistance superconducting temperatures, and there is likewise no apparent change in the widths of the transitions indicating that polishing after annealing does not introduce significant strain.
The most drastic change of polishing is that S1 now has a RRR of 30, rather than its annealed (Y) value of 20 reflecting the removal of the undue influence of the LTP. 
The RRR of S2 did not change due to polishing and remained at 20. 
The values of $\rho(T\rightarrow 0 K)$, however, did change for S1 to $\sim$ 8 $\mu \Omega$-cm and for S2 to $\sim$ 11 $\mu \Omega$-cm, reflecting the overall change in $\rho(T)$ indicated by the red and light blue arrows as was noted at higher temperatures in Fig. \ref{fig:rho} (a). 
There was no qualitative change in the shape, and hence power law behavior, in the ferromagnetic state nor the paramagnetic state above $T_c$ (not shown). 
We emphasize that the RRR of 30 for the unannealed (X) S1 sample was obtained by extrapolating to zero excitation current and that the annealed (Y) samples displayed minimal change in RRR with excitation current.
If this extrapolation was not done then the unannealed (X) data shown would also exhibit RRR$\sim$20.



\begin{figure}[h!]
	\centering
	\hspace{1em}
	\includegraphics[width=0.95\columnwidth]{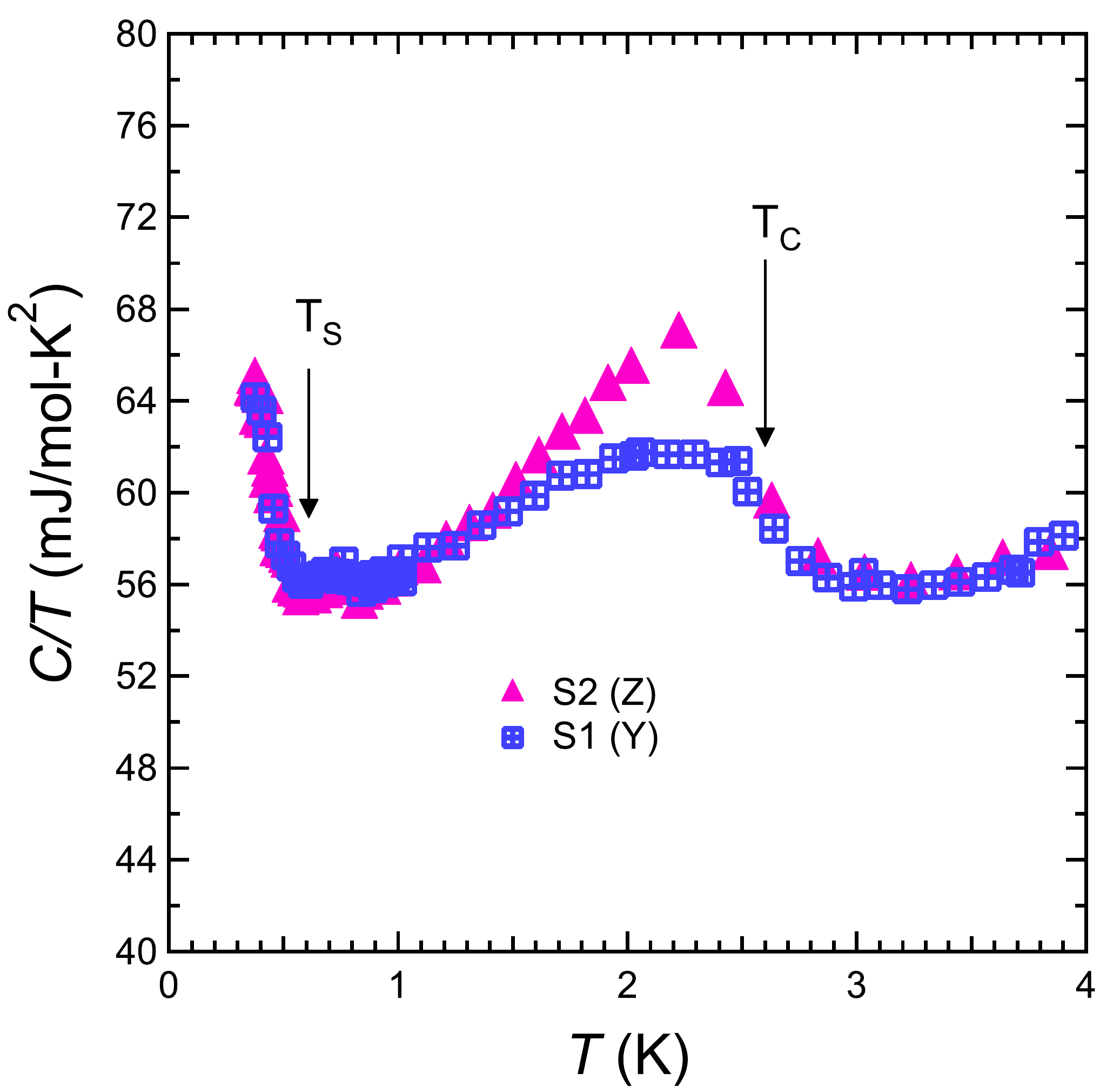}
	\caption{Heat capacity, plotted as $C/T$, vs. $T$ of zone refined UCoGe S1 annealed (Y). Anomalies at the ferromagnetic transition $T_c$ = 2.6 K and at the superconducting transition at $T_s$ = 0.61 K are observed, as indicated by the arrows. The upturn in $C/T$ provides evidence for a bulk superconducting transition.}
	\label{fig:Heat}
\end{figure}

The heat capacity, plotted as $C/T$ vs. T, from the annealed (Y) S1 sample and the annealed/polished (Z) s2 sample are presented in Fig.\,\ref{fig:Heat}. Both $T_s$ and $T_c$ as determined from $\rho$ in Fig. \,\ref{fig:rho} are indicated and show anomalies in $C/T$. The Sommerfeld coefficient can be estimated to be $\sim$ 56 mJ/mol-K$^2$ based on its magnitude to $\sim$ 3 K.
The $C(T)/T$ data show a minor inflection and broad hump with the onset of the itinerant ferromagnetic phase at $T_c$ = 2.6 K. 
Unlike a typical ferromagnetic phase transition with well localized moments, the weak itinerant ferromagnetic transition in UCoGe exhibits a minimal entropy change due to the strong hybridization of the conduction electrons with the U local moments.
This minimal entropy change at $T_c$ is relevant to the quantum critical behavior and subtle changes to the ferromagnetic and paramagnetic states observed as a function of continuous doping of UCo$_{1-x}$Rh$_x$Ge \cite{Pospisil_PRB_2020}. The ferromagnetic transition for annealed/polished (Z) S2 is sharper and has slightly more entropy change than annealed (Y) S1 sample. This is unlikely to be due to the different sample conditions (Y vs. Z) as the LTP that collected at the surface is not expected to make up a large enough mass fraction to manifest in heat capacity. If the LTP did manifest in heat capacity then the $C/T$ near $T_s$ and at temperatures above $T_c$ would not match between the two samples, which is not evident from the data. 
In contrast, $T_s$ shows the start of a noticeable rise in $C/T$ coinciding with the zero resistance superconducting temperature of 0.52 K indicating that the superconducting state occupies a significant volume of the sample, instead of existing only at the surface. The $C/T$ of both samples are indistinguishable from each other qualitatively and quantitatively in the vicinity of $T_s$. 
Comparison with the details of the heat capacity in the superconducting state between our samples and those of Refs. \cite{Aoki_JPSJ_2012, Aoki_JPSJ_2014} must await future specific heat measurements at lower temperatures than what was achieved in this study. 
There exists no indication of any behavior in $C/T$ that could be ascribed to the LTP.

It is possible that the LTP is present and undetected or unreported in previous work on UCoGe and may explain the difficulty in improving sample quality, as well as the lack of consistency in $\rho(T)$ from different reports.
It may also explain the difficulty in Shubnikov$-$de Haas (SdH) quantum oscillation measurements \cite{Aoki_JPSJ_2011, Bay_PRB_2014, Bastien_PRL_2016, Knafo_PRB_2012}.
The identity of the low temperature phase responsible for the significant change in $\rho(T)$ observed in our samples is an important issue for improving sample quality.
Due to the lack of information about the U-Co-Ge ternary system it is possible that UCoGe is a single phase that is stable over a finite composition range.
This might be reflected in the aforementioned historical disagreement in the crystal structure of UCoGe \cite{Aoki_JPSJ_2019, Gasparini_LowTempPhys_2010}. 

However, depending on the growth technique, it is likely that regions of $\sim$1:1:1 stoichiometry UCoGe make up the major volume of the crystal which hosts bulk superconductivity as shown by the observable rise in $C/T$ in Fig.\,\ref{fig:Heat}, but off-stoichiometric U(CoGe)$_{1-\delta}$ ends up stuck in cracks and voids of the crystal. It appears that the LTP is more susceptible to chemical attack by acid possibly due to the lack of Ge, which is relatively chemically inert compared to U, and Co.
In turn, the lack of Co and Ge results in a decrease of the melting temperature of U(CoGe)$_{1-\delta}$ below 900 $^\circ$C where the samples were annealed. 
There is evidence that the proper Co and Ge stoichiometry is critical in determining sample quality, with too low of a Co and Ge concentration resulting in lack of ferromagnetism and superconductivity \cite{Pospisil_JPSJ_2014}.
We plan future growth attempts with an excess of Co and Ge to compensate for the evaporation during growth that led to the formation of regions of the low temperature phase.


\begin{figure}[h!]
	\centering
	\hspace{1em}
	\includegraphics[width=0.95\columnwidth]{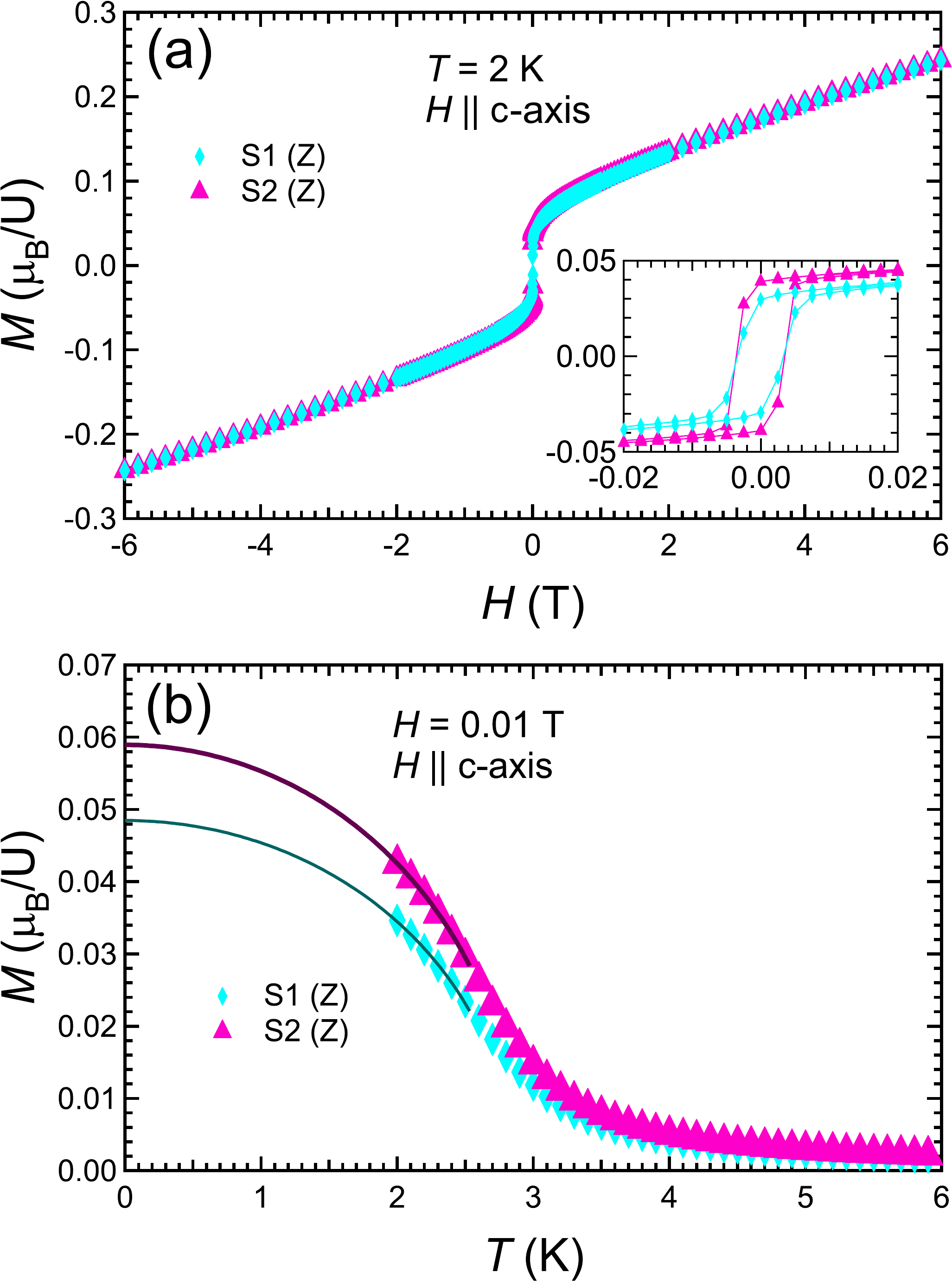}
	\caption{Magnetization, $M$, vs. field, $H$, with an inset focusing on the ferromagnetic hysteresis (a) taken at 2 K with $\bold{H}$ parallel to the c-axis. $M$ vs. $T$ upon cooling in an $H = 0.01$ T parallel to the c-axis. Solid lines are fits to $M^2(T)=M_0^2 (1-(T/T^*)^2)$ as described in the text.}
	\label{fig:Mag}
\end{figure}

The magnetization, $M$, vs. field, $H$, is shown in Fig. \ref{fig:Mag} (a) with the inset focusing on the low $H$, low $M$ region for both samples in the annealed/polished (Z) condition. The $\bold{H}$ is parallel to the Ising c-axis direction. The $|M|$ is $\sim$ 0.24 $\mu_B$ per uranium atom at the maximum $|H|$ of 6 T. At large $|H|$ both samples have nearly identical $|M|$ suggesting that both samples have the same magnetic moment per uranium atom within experimental resolution. At low fields in the inset of Fig. \ref{fig:Mag} (a) S1 has a slightly smaller $|M|$ compared to S2, but they both have nearly the same ferromagnetic coercive field of $\sim$ 0.002 T. 

The $M$ vs. $T$ is shown in Fig. \ref{fig:Mag} (b) taken upon cooling in $H$=0.01 T parallel to the c-axis. At high temperatures there is minimal spontaneous magnetic moment, but upon cooling there is a gradual onset of a spontaneous magnetic moment reflected in the increase of $M$ consistent with a ferromagnetic transition at $T_c$ of 2.5 K, as was indicated by the kink at 2.5 K in $\rho(T)$ from Fig. \ref{fig:rho} (b). The solid lines are a fit to the expression $M^2(T) = M_0^2 (1-(T/T^*)^2)$ below 2.5 K with $M_0$ and $T^*$ as free fit parameters, as was done in Refs. \cite{Huy_MagMaterials_2009, Huy_PRL_2008}. The obtained $T^*$ for both samples is $\sim$ 2.9 K, but the $M_0$ for S1 is $\sim$ 0.048 $\mu_B$ and for S2 is $\sim$ 0.059 $\mu_B$. Our $M_0$ values from fitting the $M$ vs. $T$ at low fields are similar to the previously reported values of 0.06 $\mu_B$ \cite{Huy_MagMaterials_2009} and 0.07 $\mu_B$ \cite{Huy_PRL_2008}. The mismatch in $M (T)$ between the two samples is likely due to demagnetization effects, misalignment from the c-axis, or slight difference of magnetic properties between the two samples. The $M$ behavior shown in Fig. \ref{fig:Mag} is qualitatively consistent with previous high quality samples in Refs. \cite{Huy_MagMaterials_2009, Huy_PRL_2008} and quantitatively consistent at high fields.
 
\section{Discussion}
The UCoGe samples we have prepared are of quite respectable quality with RRR $\sim$ 20-30, showing signatures of bulk superconductivity and ferromagnetism. 
However, UCoGe sample quality can not be uniquely identified with RRR as this implicitly assumes that $\rho(300 K)$ is sample quality independent, which is not self-evident from the present understanding of this material.

The sample with the highest RRR for which $\rho(T)$ was reported is by Aoki {\it et al.} \cite{Aoki_JPSJ_2014} at 110, and it shows a zero resistance temperature of $\sim$ 0.58 K with $\rho(T\rightarrow 0 K)\sim$ 3 $\mu \Omega$-cm. But in this instance the superconducting transition in the 110 RRR sample is substantially broadened suggesting poor sample homogeneity. 
Most unconventional superconductors are quite sensitive to disorder \cite{Gorkov_SovSciRevA_1987}, and exhibit suppressed transition temperatures as $\rho(T\rightarrow 0 K)$ increases such as in the case of UPt$_3$ \cite{Kycia_PRB_1998}.
Aoki {\it et al.} \cite{Aoki_JPSJ_2014} describe the evolution of $T_s$, with RRR and $\rho(T\rightarrow 0 K)$ across multiple single crystal UCoGe samples \cite{Aoki_JPSJ_2014}, and postulate that $T_s$ may be correlated linearly with 1/RRR. There is some uncertainty as to what is the best metric for sample quality for UCoGe and we suggest that $\rho(T\rightarrow 0 K)$ is a more reliable measure than 1/RRR since the high temperature resistivity appears to vary from sample to sample \cite{Hattori_PRL_2012, Aoki_JPSJ_2014, Troc_PM_2010, Pospisil_JPSJ_2014}. However, we note one inconsistent report with a zero resistance superconducting temperature of $\sim$ 0.5 K with a comparatively large $\rho(T\rightarrow 0 K)$ of 27 $\mu \Omega$-cm \cite{Troc_PM_2010}, albeit with a sample having conflicting indications of the presence/absence of bulk ferromagnetism.

Another important indication of sample quality is the consistent indication of $T_s$ from $\rho(T)$ and $C/T$ which is not always the case \cite{Huy_PRL_2007}.
Our single crystals, as well as other high quality single crystals \cite{Aoki_JPSJ_2012}, show signatures of superconductivity in both $\rho(T)$ and $C/T$ at roughly the same temperature.
A further indication of high crystal quality is a smaller $\rho(T\rightarrow 0 K)$ value. The value of $\rho(T\rightarrow 0 K)$ for S1 is $\sim$ 8 $\mu \Omega$-cm , which is low compared with the results in \cite{Aoki_JPSJ_2014, Hattori_PRL_2012, Troc_PM_2010}, and more in line with the results of \cite{Huy_PRL_2008, Huy_MagMaterials_2009}. Finally we note that with our growth method, $\rho(T)$ for unannealed (X) and annealed/polished (Z), is of the same magnitude rather than being a factor of 10 larger as reported by \cite{Huy_MagMaterials_2009}.
This may be due to the lack of large scale structural defects in our unannealed (X) samples compared to Czochralski grown crystals.
This leads to the question of why our unannealed (X) samples lack superconductivity and have a very broadened ferromagnetic transition. 
It is possible that the main effect of annealing is correcting random site occupancy of the Co and Ge via diffusion, rather than removing larger defects like stacking faults.
This would be consistent with both the aforementioned historical disagreement of the true crystal structure, and the observation of Co and Ge stoichiometry being vital for robust ferromagnetism and superconductivity \cite{Pospisil_JPSJ_2014}.

\section{Conclusion}
We have used the uncommon but effective method of UHV electron-beam zone refining to grow single crystals of the ferromagnetic superconductor UCoGe. The samples exhibit robust, bulk superconductivity and ferromagnetism, as shown by the $\rho$, $C/T$, and $M$ data, placing our samples in good company with the best UCoGe made to date. 
There is evidence of a slight loss of Co and Ge causing structural integrity issues, as well as negatively impacting the superconductivity and ferromagnetism by forming regions of off-stoichiometric U(CoGe)$_{1-\delta}$ in cracks and voids in the single crystal samples.
We have uncovered new metallurgical properties of UCoGe that may allow future improvement of UCoGe single crystal quality. 
Our ability to zone refine multiple times on a single ingot indicates that we can further improve sample quality if we properly account for the Co and Ge volatility under UHV conditions.
We have shown surface treatment to have an influence on the measured resistance of UCoGe, but not superconducting and ferromagnetic temperatures, and consequently for future work we recommend a procedure for careful polishing of the crystals to improve the interpretation of electrical transport data. 

\section{Acknowledgments}
Work at Los Alamos National Laboratory was performed under the auspices of the U.S. Department of Energy, Office of Basic
Energy Sciences, Division of Materials Sciences and Engineering. This work made use of the Jerome B. Cohen x-Ray Diffraction Facility supported by the MRSEC program of the National Science Foundation (DMR-1720139) at the Materials Research Center of Northwestern University and the Soft and Hybrid Nanotechnology Experimental (SHyNE) Resource (NSF ECCS-1542205). Support for crystal growth was provided by the U.S. Department of Energy (DOE), Office of Basic Energy Sciences (BES), Division of Material Sciences and Engineering under Award No. DE-FG02-05ER46248 (WPH).
This work made use of the EPIC facility of Northwestern University's NUANCE Center, which has received support from the SHyNE Resource (NSF ECCS-2025633), the IIN, and Northwestern's MRSEC program (NSF DMR-1720139). Support is acknowledged from the Northwestern-Fermilab Center for Applied Physics and Superconducting Technologies (CAPST). Support is also acknowledged by the U.S. Department of Energy, Office of Science, Office of Workforce Development for Teachers and Scientists, Office of Science Graduate Student Research (SCGSR) program. The SCGSR program is administered by the Oak Ridge Institute for Science and Education for the DOE under contract number DE-SC0014664.




\begin{thebibliography}{57}%
\makeatletter
\providecommand \@ifxundefined [1]{%
 \@ifx{#1\undefined}
}%
\providecommand \@ifnum [1]{%
 \ifnum #1\expandafter \@firstoftwo
 \else \expandafter \@secondoftwo
 \fi
}%
\providecommand \@ifx [1]{%
 \ifx #1\expandafter \@firstoftwo
 \else \expandafter \@secondoftwo
 \fi
}%
\providecommand \natexlab [1]{#1}%
\providecommand \enquote  [1]{``#1''}%
\providecommand \bibnamefont  [1]{#1}%
\providecommand \bibfnamefont [1]{#1}%
\providecommand \citenamefont [1]{#1}%
\providecommand \href@noop [0]{\@secondoftwo}%
\providecommand \href [0]{\begingroup \@sanitize@url \@href}%
\providecommand \@href[1]{\@@startlink{#1}\@@href}%
\providecommand \@@href[1]{\endgroup#1\@@endlink}%
\providecommand \@sanitize@url [0]{\catcode `\\12\catcode `\$12\catcode
  `\&12\catcode `\#12\catcode `\^12\catcode `\_12\catcode `\%12\relax}%
\providecommand \@@startlink[1]{}%
\providecommand \@@endlink[0]{}%
\providecommand \url  [0]{\begingroup\@sanitize@url \@url }%
\providecommand \@url [1]{\endgroup\@href {#1}{\urlprefix }}%
\providecommand \urlprefix  [0]{URL }%
\providecommand \Eprint [0]{\href }%
\providecommand \doibase [0]{https://doi.org/}%
\providecommand \selectlanguage [0]{\@gobble}%
\providecommand \bibinfo  [0]{\@secondoftwo}%
\providecommand \bibfield  [0]{\@secondoftwo}%
\providecommand \translation [1]{[#1]}%
\providecommand \BibitemOpen [0]{}%
\providecommand \bibitemStop [0]{}%
\providecommand \bibitemNoStop [0]{.\EOS\space}%
\providecommand \EOS [0]{\spacefactor3000\relax}%
\providecommand \BibitemShut  [1]{\csname bibitem#1\endcsname}%
\let\auto@bib@innerbib\@empty
\bibitem [{\citenamefont {Onnes}(1911)}]{Onnes_Proceedings_1911}%
  \BibitemOpen
  \bibfield  {author} {\bibinfo {author} {\bibfnamefont {H.~K.}\ \bibnamefont
  {Onnes}},\ }\href@noop {} {\bibfield  {journal} {\bibinfo  {journal}
  {Proceedings}\ }\textbf {\bibinfo {volume} {13 II}},\ \bibinfo {pages} {1274}
  (\bibinfo {year} {1911})}\BibitemShut {NoStop}%
\bibitem [{\citenamefont {van Delft}\ and\ \citenamefont
  {Kes}(2011)}]{Delft_EuroPhysNews_2011}%
  \BibitemOpen
  \bibfield  {author} {\bibinfo {author} {\bibfnamefont {D.}~\bibnamefont {van
  Delft}}\ and\ \bibinfo {author} {\bibfnamefont {P.}~\bibnamefont {Kes}},\
  }\bibfield  {title} {\bibinfo {title} {The {D}iscovery of
  {S}uperconductivity},\ }\href@noop {} {\bibfield  {journal} {\bibinfo
  {journal} {Europhysics News}\ }\textbf {\bibinfo {volume} {42}},\ \bibinfo
  {pages} {21} (\bibinfo {year} {2011})}\BibitemShut {NoStop}%
\bibitem [{\citenamefont {Bardeen}\ \emph {et~al.}(1957)\citenamefont
  {Bardeen}, \citenamefont {Cooper},\ and\ \citenamefont
  {Schrieffer}}]{Bardeen_PhysicalReview_1957}%
  \BibitemOpen
  \bibfield  {author} {\bibinfo {author} {\bibfnamefont {J.}~\bibnamefont
  {Bardeen}}, \bibinfo {author} {\bibfnamefont {L.~N.}\ \bibnamefont
  {Cooper}},\ and\ \bibinfo {author} {\bibfnamefont {J.~R.}\ \bibnamefont
  {Schrieffer}},\ }\bibfield  {title} {\bibinfo {title} {Theory of
  {S}uperconductivity},\ }\href@noop {} {\bibfield  {journal} {\bibinfo
  {journal} {Phys. Rev.}\ }\textbf {\bibinfo {volume} {108}},\ \bibinfo {pages}
  {1175} (\bibinfo {year} {1957})}\BibitemShut {NoStop}%
\bibitem [{\citenamefont {Bednorz}\ and\ \citenamefont
  {Muller}(1986)}]{Bednorz_ZPB_1986}%
  \BibitemOpen
  \bibfield  {author} {\bibinfo {author} {\bibfnamefont {J.~G.}\ \bibnamefont
  {Bednorz}}\ and\ \bibinfo {author} {\bibfnamefont {K.~A.}\ \bibnamefont
  {Muller}},\ }\bibfield  {title} {\bibinfo {title} {Possible {H}igh {T}$_c$
  {S}uperconductivity in the {B}a-{L}a-{C}u-{O} {S}ystem},\ }\href@noop {}
  {\bibfield  {journal} {\bibinfo  {journal} {Z. Phys. B}\ }\textbf {\bibinfo
  {volume} {64}},\ \bibinfo {pages} {189} (\bibinfo {year} {1986})}\BibitemShut
  {NoStop}%
\bibitem [{\citenamefont {Wu}\ \emph {et~al.}(1987)\citenamefont {Wu},
  \citenamefont {Ashburn}, \citenamefont {Torng}, \citenamefont {Hor},
  \citenamefont {Meng}, \citenamefont {Gao}, \citenamefont {Huang},
  \citenamefont {Wang},\ and\ \citenamefont {Chu}}]{Wu_PRL_1987}%
  \BibitemOpen
  \bibfield  {author} {\bibinfo {author} {\bibfnamefont {M.~K.}\ \bibnamefont
  {Wu}}, \bibinfo {author} {\bibfnamefont {J.~R.}\ \bibnamefont {Ashburn}},
  \bibinfo {author} {\bibfnamefont {C.~J.}\ \bibnamefont {Torng}}, \bibinfo
  {author} {\bibfnamefont {P.~H.}\ \bibnamefont {Hor}}, \bibinfo {author}
  {\bibfnamefont {R.~L.}\ \bibnamefont {Meng}}, \bibinfo {author}
  {\bibfnamefont {L.}~\bibnamefont {Gao}}, \bibinfo {author} {\bibfnamefont
  {Z.~J.}\ \bibnamefont {Huang}}, \bibinfo {author} {\bibfnamefont {Y.~Q.}\
  \bibnamefont {Wang}},\ and\ \bibinfo {author} {\bibfnamefont {C.~W.}\
  \bibnamefont {Chu}},\ }\bibfield  {title} {\bibinfo {title}
  {Superconductivity at 93 {K} in a {N}ew {M}ixed-{P}hase {Y}-{B}a-{C}u-{O}
  {C}ompound {S}ystem at {A}mbient {P}ressure},\ }\href@noop {} {\bibfield
  {journal} {\bibinfo  {journal} {Phys. Rev. Lett.}\ }\textbf {\bibinfo
  {volume} {58}},\ \bibinfo {pages} {908} (\bibinfo {year} {1987})}\BibitemShut
  {NoStop}%
\bibitem [{\citenamefont {Steglich}\ \emph {et~al.}(1979)\citenamefont
  {Steglich}, \citenamefont {Aarts}, \citenamefont {Bredl}, \citenamefont
  {Lieke}, \citenamefont {Meschede}, \citenamefont {Franz},\ and\ \citenamefont
  {Schafer}}]{Steglich_PRL_1979}%
  \BibitemOpen
  \bibfield  {author} {\bibinfo {author} {\bibfnamefont {F.}~\bibnamefont
  {Steglich}}, \bibinfo {author} {\bibfnamefont {J.}~\bibnamefont {Aarts}},
  \bibinfo {author} {\bibfnamefont {C.~D.}\ \bibnamefont {Bredl}}, \bibinfo
  {author} {\bibfnamefont {W.}~\bibnamefont {Lieke}}, \bibinfo {author}
  {\bibfnamefont {D.}~\bibnamefont {Meschede}}, \bibinfo {author}
  {\bibfnamefont {W.}~\bibnamefont {Franz}},\ and\ \bibinfo {author}
  {\bibfnamefont {H.}~\bibnamefont {Schafer}},\ }\bibfield  {title} {\bibinfo
  {title} {Superconductivity in the {P}resence of {S}trong {P}auli
  {P}aramagnetism: {C}e{C}u$_2${S}i$_2$},\ }\href@noop {} {\bibfield  {journal}
  {\bibinfo  {journal} {Phys. Rev. Lett.}\ }\textbf {\bibinfo {volume} {43}},\
  \bibinfo {pages} {1892} (\bibinfo {year} {1979})}\BibitemShut {NoStop}%
\bibitem [{\citenamefont {Schemm}\ \emph {et~al.}(2014)\citenamefont {Schemm},
  \citenamefont {Gannon}, \citenamefont {Wishne}, \citenamefont {Halperin},\
  and\ \citenamefont {Kapitulnik}}]{Schemm_Science_2014}%
  \BibitemOpen
  \bibfield  {author} {\bibinfo {author} {\bibfnamefont {E.~R.}\ \bibnamefont
  {Schemm}}, \bibinfo {author} {\bibfnamefont {W.~J.}\ \bibnamefont {Gannon}},
  \bibinfo {author} {\bibfnamefont {C.~M.}\ \bibnamefont {Wishne}}, \bibinfo
  {author} {\bibfnamefont {W.~P.}\ \bibnamefont {Halperin}},\ and\ \bibinfo
  {author} {\bibfnamefont {A.}~\bibnamefont {Kapitulnik}},\ }\bibfield  {title}
  {\bibinfo {title} {Observation of {B}roken {T}ime-{R}eversal {S}ymmetry in
  the {H}eavy-{F}ermion {S}uperconductor {U}{P}t$_{3}$},\ }\href@noop {}
  {\bibfield  {journal} {\bibinfo  {journal} {Science}\ }\textbf {\bibinfo
  {volume} {345}},\ \bibinfo {pages} {190} (\bibinfo {year}
  {2014})}\BibitemShut {NoStop}%
\bibitem [{\citenamefont {Avers}\ \emph {et~al.}(2020)\citenamefont {Avers},
  \citenamefont {Gannon}, \citenamefont {Kuhn}, \citenamefont {Halperin},
  \citenamefont {Sauls}, \citenamefont {DeBeer-Schmitt}, \citenamefont
  {Dewhurst}, \citenamefont {Gavilano}, \citenamefont {Nagy}, \citenamefont
  {Gasser},\ and\ \citenamefont {Eskildsen}}]{Avers_NaturePhys_2020}%
  \BibitemOpen
  \bibfield  {author} {\bibinfo {author} {\bibfnamefont {K.~E.}\ \bibnamefont
  {Avers}}, \bibinfo {author} {\bibfnamefont {W.~J.}\ \bibnamefont {Gannon}},
  \bibinfo {author} {\bibfnamefont {S.~J.}\ \bibnamefont {Kuhn}}, \bibinfo
  {author} {\bibfnamefont {W.~P.}\ \bibnamefont {Halperin}}, \bibinfo {author}
  {\bibfnamefont {J.~A.}\ \bibnamefont {Sauls}}, \bibinfo {author}
  {\bibfnamefont {L.}~\bibnamefont {DeBeer-Schmitt}}, \bibinfo {author}
  {\bibfnamefont {C.~D.}\ \bibnamefont {Dewhurst}}, \bibinfo {author}
  {\bibfnamefont {J.}~\bibnamefont {Gavilano}}, \bibinfo {author}
  {\bibfnamefont {G.}~\bibnamefont {Nagy}}, \bibinfo {author} {\bibfnamefont
  {U.}~\bibnamefont {Gasser}},\ and\ \bibinfo {author} {\bibfnamefont {M.~R.}\
  \bibnamefont {Eskildsen}},\ }\bibfield  {title} {\bibinfo {title} {Broken
  {T}ime-{R}eversal {S}ymmetry {B}reaking in the {T}opological {S}uperconductor
  {U}{P}t$_3$},\ }\bibfield  {journal} {\bibinfo  {journal} {Nature Phys.}\
  }\href {https://doi.org/https://doi.org/10.1038/s41567-020-0822-z}
  {https://doi.org/10.1038/s41567-020-0822-z} (\bibinfo {year}
  {2020})\BibitemShut {NoStop}%
\bibitem [{\citenamefont {Wartenbe}\ \emph {et~al.}(2019)\citenamefont
  {Wartenbe}, \citenamefont {Baumbach}, \citenamefont {Shekhter}, \citenamefont
  {Boebinger}, \citenamefont {Bauer}, \citenamefont {Moya}, \citenamefont
  {Harrison}, \citenamefont {McDonald}, \citenamefont {Salamon},\ and\
  \citenamefont {Jamie}}]{Wartenbe_PRB_2019}%
  \BibitemOpen
  \bibfield  {author} {\bibinfo {author} {\bibfnamefont {M.}~\bibnamefont
  {Wartenbe}}, \bibinfo {author} {\bibfnamefont {R.~E.}\ \bibnamefont
  {Baumbach}}, \bibinfo {author} {\bibfnamefont {A.}~\bibnamefont {Shekhter}},
  \bibinfo {author} {\bibfnamefont {G.~S.}\ \bibnamefont {Boebinger}}, \bibinfo
  {author} {\bibfnamefont {E.~D.}\ \bibnamefont {Bauer}}, \bibinfo {author}
  {\bibfnamefont {C.~C.}\ \bibnamefont {Moya}}, \bibinfo {author}
  {\bibfnamefont {N.}~\bibnamefont {Harrison}}, \bibinfo {author}
  {\bibfnamefont {R.~D.}\ \bibnamefont {McDonald}}, \bibinfo {author}
  {\bibfnamefont {M.~B.}\ \bibnamefont {Salamon}},\ and\ \bibinfo {author}
  {\bibfnamefont {M.}~\bibnamefont {Jamie}},\ }\bibfield  {title} {\bibinfo
  {title} {Magnetoelastic {C}oupling in {U}{R}u$_2${S}i$_2$: {P}robing
  {M}ultipolar {C}orrelations in the {H}idden {O}rder {S}tate},\ }\href@noop {}
  {\bibfield  {journal} {\bibinfo  {journal} {Phys. Rev. B}\ }\textbf {\bibinfo
  {volume} {99}},\ \bibinfo {pages} {235101} (\bibinfo {year}
  {2019})}\BibitemShut {NoStop}%
\bibitem [{\citenamefont {Schemm}\ \emph {et~al.}(2015)\citenamefont {Schemm},
  \citenamefont {Baumbach}, \citenamefont {Tobash}, \citenamefont {Ronning},
  \citenamefont {Bauer},\ and\ \citenamefont {Kapitulnik}}]{Schemm_PRB_2015}%
  \BibitemOpen
  \bibfield  {author} {\bibinfo {author} {\bibfnamefont {E.~R.}\ \bibnamefont
  {Schemm}}, \bibinfo {author} {\bibfnamefont {R.~E.}\ \bibnamefont
  {Baumbach}}, \bibinfo {author} {\bibfnamefont {P.~H.}\ \bibnamefont
  {Tobash}}, \bibinfo {author} {\bibfnamefont {F.}~\bibnamefont {Ronning}},
  \bibinfo {author} {\bibfnamefont {E.~D.}\ \bibnamefont {Bauer}},\ and\
  \bibinfo {author} {\bibfnamefont {A.}~\bibnamefont {Kapitulnik}},\ }\bibfield
   {title} {\bibinfo {title} {Evidence for {B}roken {T}ime-{R}eversal
  {S}ymmetry in the {S}upercondcuting {P}hase of {U}{R}u$_2${S}i$_2$},\
  }\href@noop {} {\bibfield  {journal} {\bibinfo  {journal} {Phys. Rev. B}\
  }\textbf {\bibinfo {volume} {91}},\ \bibinfo {pages} {140506(R)} (\bibinfo
  {year} {2015})}\BibitemShut {NoStop}%
\bibitem [{\citenamefont {Adenwalla}\ \emph {et~al.}(1990)\citenamefont
  {Adenwalla}, \citenamefont {Lin}, \citenamefont {Ran}, \citenamefont {Zhao},
  \citenamefont {Ketterson}, \citenamefont {Sauls}, \citenamefont {Taillefer},
  \citenamefont {Hinks}, \citenamefont {Levy},\ and\ \citenamefont
  {Sarma}}]{Adenwalla_PRL_1990}%
  \BibitemOpen
  \bibfield  {author} {\bibinfo {author} {\bibfnamefont {S.}~\bibnamefont
  {Adenwalla}}, \bibinfo {author} {\bibfnamefont {S.~W.}\ \bibnamefont {Lin}},
  \bibinfo {author} {\bibfnamefont {Q.~Z.}\ \bibnamefont {Ran}}, \bibinfo
  {author} {\bibfnamefont {Z.}~\bibnamefont {Zhao}}, \bibinfo {author}
  {\bibfnamefont {J.~B.}\ \bibnamefont {Ketterson}}, \bibinfo {author}
  {\bibfnamefont {J.~A.}\ \bibnamefont {Sauls}}, \bibinfo {author}
  {\bibfnamefont {L.}~\bibnamefont {Taillefer}}, \bibinfo {author}
  {\bibfnamefont {D.~G.}\ \bibnamefont {Hinks}}, \bibinfo {author}
  {\bibfnamefont {M.}~\bibnamefont {Levy}},\ and\ \bibinfo {author}
  {\bibfnamefont {B.~K.}\ \bibnamefont {Sarma}},\ }\bibfield  {title} {\bibinfo
  {title} {Phase {D}iagram of {U}{P}t$_3$ from {U}ltrasonic {V}elocity
  {M}easurements},\ }\href@noop {} {\bibfield  {journal} {\bibinfo  {journal}
  {Phys. Rev. Lett.}\ }\textbf {\bibinfo {volume} {65}},\ \bibinfo {pages}
  {2298} (\bibinfo {year} {1990})}\BibitemShut {NoStop}%
\bibitem [{\citenamefont {Stewart}(2019)}]{Stewart_LowTempPhys_2019}%
  \BibitemOpen
  \bibfield  {author} {\bibinfo {author} {\bibfnamefont {G.~R.}\ \bibnamefont
  {Stewart}},\ }\bibfield  {title} {\bibinfo {title} {U{B}e$_{13}$ and
  {U}$_{1-x}${T}h$_{x}${B}e$_{13}$:{U}nconventional {S}uperconductors},\
  }\href@noop {} {\bibfield  {journal} {\bibinfo  {journal} {J. of Low Temp.
  Phys.}\ }\textbf {\bibinfo {volume} {195}},\ \bibinfo {pages} {1} (\bibinfo
  {year} {2019})}\BibitemShut {NoStop}%
\bibitem [{\citenamefont {Werthamer}\ \emph {et~al.}(1966)\citenamefont
  {Werthamer}, \citenamefont {Helfand},\ and\ \citenamefont
  {Hoenberg}}]{Werthamer_PRL_1966}%
  \BibitemOpen
  \bibfield  {author} {\bibinfo {author} {\bibfnamefont {N.~R.}\ \bibnamefont
  {Werthamer}}, \bibinfo {author} {\bibfnamefont {E.}~\bibnamefont {Helfand}},\
  and\ \bibinfo {author} {\bibfnamefont {P.~C.}\ \bibnamefont {Hoenberg}},\
  }\bibfield  {title} {\bibinfo {title} {Temperature and {P}urity {D}ependence
  of the {S}uperconducting {C}ritical {F}ield, {H}$_{c2}$. {I}{I}{I}.
  {E}lectron {S}pin and {S}pin-{O}rbit {E}ffects},\ }\href@noop {} {\bibfield
  {journal} {\bibinfo  {journal} {Phys. Rev.}\ }\textbf {\bibinfo {volume}
  {147}},\ \bibinfo {pages} {295} (\bibinfo {year} {1966})}\BibitemShut
  {NoStop}%
\bibitem [{\citenamefont {Suhl}\ \emph {et~al.}(1959)\citenamefont {Suhl},
  \citenamefont {Matthias},\ and\ \citenamefont
  {Corenzwit}}]{Suhl_PhysAndChemOfSolids_1959}%
  \BibitemOpen
  \bibfield  {author} {\bibinfo {author} {\bibfnamefont {H.}~\bibnamefont
  {Suhl}}, \bibinfo {author} {\bibfnamefont {B.~T.}\ \bibnamefont {Matthias}},\
  and\ \bibinfo {author} {\bibfnamefont {E.}~\bibnamefont {Corenzwit}},\
  }\bibfield  {title} {\bibinfo {title} {Some {F}urther {R}esults on
  {F}erromagnetism in {R}elation to {S}uperconductivity},\ }\href@noop {}
  {\bibfield  {journal} {\bibinfo  {journal} {J. Phys. Chem. Solid}\ }\textbf
  {\bibinfo {volume} {11}},\ \bibinfo {pages} {346} (\bibinfo {year}
  {1959})}\BibitemShut {NoStop}%
\bibitem [{\citenamefont {Matthias}\ \emph {et~al.}(1958)\citenamefont
  {Matthias}, \citenamefont {Suhl},\ and\ \citenamefont
  {Corenzwit}}]{Matthias_PRL_1958}%
  \BibitemOpen
  \bibfield  {author} {\bibinfo {author} {\bibfnamefont {B.~T.}\ \bibnamefont
  {Matthias}}, \bibinfo {author} {\bibfnamefont {H.}~\bibnamefont {Suhl}},\
  and\ \bibinfo {author} {\bibfnamefont {E.}~\bibnamefont {Corenzwit}},\
  }\bibfield  {title} {\bibinfo {title} {Ferromagnetic {S}uperconductors},\
  }\href@noop {} {\bibfield  {journal} {\bibinfo  {journal} {Phys. Rev. Lett.}\
  }\textbf {\bibinfo {volume} {1}},\ \bibinfo {pages} {449} (\bibinfo {year}
  {1958})}\BibitemShut {NoStop}%
\bibitem [{\citenamefont {Liu}\ \emph {et~al.}(2017)\citenamefont {Liu},
  \citenamefont {Liu}, \citenamefont {Yu}, \citenamefont {Tao}, \citenamefont
  {Feng},\ and\ \citenamefont {Cao}}]{Liu_PRB_2017}%
  \BibitemOpen
  \bibfield  {author} {\bibinfo {author} {\bibfnamefont {Y.}~\bibnamefont
  {Liu}}, \bibinfo {author} {\bibfnamefont {Y.-B.}\ \bibnamefont {Liu}},
  \bibinfo {author} {\bibfnamefont {Y.-L.}\ \bibnamefont {Yu}}, \bibinfo
  {author} {\bibfnamefont {Q.}~\bibnamefont {Tao}}, \bibinfo {author}
  {\bibfnamefont {C.-M.}\ \bibnamefont {Feng}},\ and\ \bibinfo {author}
  {\bibfnamefont {G.-H.}\ \bibnamefont {Cao}},\ }\bibfield  {title} {\bibinfo
  {title} {Rb{E}u({F}e$_{1-x}${N}i$_x$)$_4${A}s$_4$: {F}rom a {F}erromagnetic
  {S}uperconductor to a {S}uperconducting {F}erromagnet},\ }\href@noop {}
  {\bibfield  {journal} {\bibinfo  {journal} {Phys. Rev. B}\ }\textbf {\bibinfo
  {volume} {96}},\ \bibinfo {pages} {224510} (\bibinfo {year}
  {2017})}\BibitemShut {NoStop}%
\bibitem [{\citenamefont {Saxena}\ \emph {et~al.}(2000)\citenamefont {Saxena},
  \citenamefont {Agarwal}, \citenamefont {Ahilan}, \citenamefont {Grosche},
  \citenamefont {Haselwimmer}, \citenamefont {Steiner}, \citenamefont {Pugh},
  \citenamefont {Walker}, \citenamefont {Julian}, \citenamefont {Monthoux}
  \emph {et~al.}}]{Saxena_Nature_2000}%
  \BibitemOpen
  \bibfield  {author} {\bibinfo {author} {\bibfnamefont {S.~S.}\ \bibnamefont
  {Saxena}}, \bibinfo {author} {\bibfnamefont {P.}~\bibnamefont {Agarwal}},
  \bibinfo {author} {\bibfnamefont {K.}~\bibnamefont {Ahilan}}, \bibinfo
  {author} {\bibfnamefont {F.~M.}\ \bibnamefont {Grosche}}, \bibinfo {author}
  {\bibfnamefont {R.~K.~W.}\ \bibnamefont {Haselwimmer}}, \bibinfo {author}
  {\bibfnamefont {M.~J.}\ \bibnamefont {Steiner}}, \bibinfo {author}
  {\bibfnamefont {E.}~\bibnamefont {Pugh}}, \bibinfo {author} {\bibfnamefont
  {I.~R.}\ \bibnamefont {Walker}}, \bibinfo {author} {\bibfnamefont {S.~R.}\
  \bibnamefont {Julian}}, \bibinfo {author} {\bibfnamefont {P.}~\bibnamefont
  {Monthoux}}, \emph {et~al.},\ }\bibfield  {title} {\bibinfo {title}
  {Superconductivity on the {B}order of {I}tinerant-{E}lectron {F}erromagnetism
  in {U}{G}e$_2$},\ }\href@noop {} {\bibfield  {journal} {\bibinfo  {journal}
  {Nature}\ }\textbf {\bibinfo {volume} {406}},\ \bibinfo {pages} {587}
  (\bibinfo {year} {2000})}\BibitemShut {NoStop}%
\bibitem [{\citenamefont {Huy}\ \emph {et~al.}(2008)\citenamefont {Huy},
  \citenamefont {de~Nijs}, \citenamefont {Huang},\ and\ \citenamefont
  {deVisser}}]{Huy_PRL_2008}%
  \BibitemOpen
  \bibfield  {author} {\bibinfo {author} {\bibfnamefont {N.~T.}\ \bibnamefont
  {Huy}}, \bibinfo {author} {\bibfnamefont {D.~E.}\ \bibnamefont {de~Nijs}},
  \bibinfo {author} {\bibfnamefont {Y.~K.}\ \bibnamefont {Huang}},\ and\
  \bibinfo {author} {\bibfnamefont {A.}~\bibnamefont {deVisser}},\ }\bibfield
  {title} {\bibinfo {title} {Unusual {U}pper {C}ritical {F}ield of the
  {F}erromagnetic {S}uperconductor {U}{C}o{G}e},\ }\href@noop {} {\bibfield
  {journal} {\bibinfo  {journal} {Phys. Rev. Lett.}\ }\textbf {\bibinfo
  {volume} {100}},\ \bibinfo {pages} {077002} (\bibinfo {year}
  {2008})}\BibitemShut {NoStop}%
\bibitem [{\citenamefont {Aoki}\ \emph {et~al.}(2001)\citenamefont {Aoki},
  \citenamefont {Huxley}, \citenamefont {Ressouche}, \citenamefont
  {Braithwaite}, \citenamefont {Flouquet}, \citenamefont {Brison},
  \citenamefont {Lhotel},\ and\ \citenamefont {Paulsen}}]{Aoki_Nature_2001}%
  \BibitemOpen
  \bibfield  {author} {\bibinfo {author} {\bibfnamefont {D.}~\bibnamefont
  {Aoki}}, \bibinfo {author} {\bibfnamefont {A.}~\bibnamefont {Huxley}},
  \bibinfo {author} {\bibfnamefont {E.}~\bibnamefont {Ressouche}}, \bibinfo
  {author} {\bibfnamefont {D.}~\bibnamefont {Braithwaite}}, \bibinfo {author}
  {\bibfnamefont {J.}~\bibnamefont {Flouquet}}, \bibinfo {author}
  {\bibfnamefont {J.-P.}\ \bibnamefont {Brison}}, \bibinfo {author}
  {\bibfnamefont {E.}~\bibnamefont {Lhotel}},\ and\ \bibinfo {author}
  {\bibfnamefont {C.}~\bibnamefont {Paulsen}},\ }\bibfield  {title} {\bibinfo
  {title} {Coexistence of {S}uperconductivity and {F}erromagnetism in
  {U}{R}h{G}e},\ }\href@noop {} {\bibfield  {journal} {\bibinfo  {journal}
  {Nature}\ }\textbf {\bibinfo {volume} {413}},\ \bibinfo {pages} {613}
  (\bibinfo {year} {2001})}\BibitemShut {NoStop}%
\bibitem [{\citenamefont {Ran}\ \emph {et~al.}(2019)\citenamefont {Ran},
  \citenamefont {Eckberg}, \citenamefont {Ding}, \citenamefont {Furukawa},
  \citenamefont {Metz}, \citenamefont {Saha}, \citenamefont {Liu},
  \citenamefont {Zic}, \citenamefont {Kim}, \citenamefont {Paglione},\ and\
  \citenamefont {Butch}}]{Ran_Science_2019}%
  \BibitemOpen
  \bibfield  {author} {\bibinfo {author} {\bibfnamefont {S.}~\bibnamefont
  {Ran}}, \bibinfo {author} {\bibfnamefont {C.}~\bibnamefont {Eckberg}},
  \bibinfo {author} {\bibfnamefont {Q.-P.}\ \bibnamefont {Ding}}, \bibinfo
  {author} {\bibfnamefont {Y.}~\bibnamefont {Furukawa}}, \bibinfo {author}
  {\bibfnamefont {T.}~\bibnamefont {Metz}}, \bibinfo {author} {\bibfnamefont
  {S.~R.}\ \bibnamefont {Saha}}, \bibinfo {author} {\bibfnamefont {I.-L.}\
  \bibnamefont {Liu}}, \bibinfo {author} {\bibfnamefont {M.}~\bibnamefont
  {Zic}}, \bibinfo {author} {\bibfnamefont {H.}~\bibnamefont {Kim}}, \bibinfo
  {author} {\bibfnamefont {J.}~\bibnamefont {Paglione}},\ and\ \bibinfo
  {author} {\bibfnamefont {N.~P.}\ \bibnamefont {Butch}},\ }\bibfield  {title}
  {\bibinfo {title} {Nearly {F}erromagnetic {S}pin-{T}riplet
  {S}uperconductivity},\ }\href@noop {} {\bibfield  {journal} {\bibinfo
  {journal} {Science}\ }\textbf {\bibinfo {volume} {365}},\ \bibinfo {pages}
  {684} (\bibinfo {year} {2019})}\BibitemShut {NoStop}%
\bibitem [{\citenamefont {Aoki}\ \emph
  {et~al.}(2019{\natexlab{a}})\citenamefont {Aoki}, \citenamefont {Nakamura},
  \citenamefont {Honda}, \citenamefont {Li}, \citenamefont {Homma},
  \citenamefont {Shimizu} \emph {et~al.}}]{Aoki2_JPSJ_2019}%
  \BibitemOpen
  \bibfield  {author} {\bibinfo {author} {\bibfnamefont {D.}~\bibnamefont
  {Aoki}}, \bibinfo {author} {\bibfnamefont {A.}~\bibnamefont {Nakamura}},
  \bibinfo {author} {\bibfnamefont {F.}~\bibnamefont {Honda}}, \bibinfo
  {author} {\bibfnamefont {D.}~\bibnamefont {Li}}, \bibinfo {author}
  {\bibfnamefont {Y.}~\bibnamefont {Homma}}, \bibinfo {author} {\bibfnamefont
  {Y.}~\bibnamefont {Shimizu}}, \emph {et~al.},\ }\bibfield  {title} {\bibinfo
  {title} {Unconventional {S}uperconductivity in {H}eavy {F}ermion
  {U}{T}e$_2$},\ }\href@noop {} {\bibfield  {journal} {\bibinfo  {journal} {J.
  Phys. Soc. Japan}\ }\textbf {\bibinfo {volume} {88}},\ \bibinfo {pages}
  {043702} (\bibinfo {year} {2019}{\natexlab{a}})}\BibitemShut {NoStop}%
\bibitem [{\citenamefont {Wu}\ \emph {et~al.}(2017)\citenamefont {Wu},
  \citenamefont {Bastien}, \citenamefont {Taupin}, \citenamefont {Paulsen},
  \citenamefont {Howald}, \citenamefont {Aoki},\ and\ \citenamefont
  {Brison}}]{Wu_NatComm_2017}%
  \BibitemOpen
  \bibfield  {author} {\bibinfo {author} {\bibfnamefont {B.}~\bibnamefont
  {Wu}}, \bibinfo {author} {\bibfnamefont {G.}~\bibnamefont {Bastien}},
  \bibinfo {author} {\bibfnamefont {M.}~\bibnamefont {Taupin}}, \bibinfo
  {author} {\bibfnamefont {C.}~\bibnamefont {Paulsen}}, \bibinfo {author}
  {\bibfnamefont {L.}~\bibnamefont {Howald}}, \bibinfo {author} {\bibfnamefont
  {D.}~\bibnamefont {Aoki}},\ and\ \bibinfo {author} {\bibfnamefont {J.-P.}\
  \bibnamefont {Brison}},\ }\bibfield  {title} {\bibinfo {title} {Pairing
  {M}echanism in the {F}erromagnetic {S}uperconductor {U}{C}o{G}e},\
  }\href@noop {} {\bibfield  {journal} {\bibinfo  {journal} {Nature Commun.}\
  }\textbf {\bibinfo {volume} {8}},\ \bibinfo {pages} {14480} (\bibinfo {year}
  {2017})}\BibitemShut {NoStop}%
\bibitem [{\citenamefont {Sundar}\ \emph {et~al.}(2019)\citenamefont {Sundar},
  \citenamefont {Gheidi}, \citenamefont {Akintola}, \citenamefont {Cote},
  \citenamefont {Dunsiger}, \citenamefont {Ran}, \citenamefont {Butch},
  \citenamefont {Saha}, \citenamefont {Paglione},\ and\ \citenamefont
  {Sonier}}]{Sundar_PRL_2019}%
  \BibitemOpen
  \bibfield  {author} {\bibinfo {author} {\bibfnamefont {S.}~\bibnamefont
  {Sundar}}, \bibinfo {author} {\bibfnamefont {S.}~\bibnamefont {Gheidi}},
  \bibinfo {author} {\bibfnamefont {K.}~\bibnamefont {Akintola}}, \bibinfo
  {author} {\bibfnamefont {A.~M.}\ \bibnamefont {Cote}}, \bibinfo {author}
  {\bibfnamefont {S.~R.}\ \bibnamefont {Dunsiger}}, \bibinfo {author}
  {\bibfnamefont {S.}~\bibnamefont {Ran}}, \bibinfo {author} {\bibfnamefont
  {N.~P.}\ \bibnamefont {Butch}}, \bibinfo {author} {\bibfnamefont {S.~R.}\
  \bibnamefont {Saha}}, \bibinfo {author} {\bibfnamefont {J.}~\bibnamefont
  {Paglione}},\ and\ \bibinfo {author} {\bibfnamefont {J.~E.}\ \bibnamefont
  {Sonier}},\ }\bibfield  {title} {\bibinfo {title} {Coexistence of
  {F}erromagnetic {F}luctuations and {S}uperconductivity in the {A}ctinide
  {S}uperconductor {U}{T}e$_2$},\ }\href@noop {} {\bibfield  {journal}
  {\bibinfo  {journal} {Phys. Rev. B}\ }\textbf {\bibinfo {volume} {100}},\
  \bibinfo {pages} {140502(R)} (\bibinfo {year} {2019})}\BibitemShut {NoStop}%
\bibitem [{\citenamefont {Tokunaga}\ \emph {et~al.}(2019)\citenamefont
  {Tokunaga}, \citenamefont {Sakai}, \citenamefont {Kambe}, \citenamefont
  {Hattori}, \citenamefont {Higa}, \citenamefont {Nakamine} \emph
  {et~al.}}]{Tokunaga_JPSJ_2019}%
  \BibitemOpen
  \bibfield  {author} {\bibinfo {author} {\bibfnamefont {Y.}~\bibnamefont
  {Tokunaga}}, \bibinfo {author} {\bibfnamefont {H.}~\bibnamefont {Sakai}},
  \bibinfo {author} {\bibfnamefont {S.}~\bibnamefont {Kambe}}, \bibinfo
  {author} {\bibfnamefont {T.}~\bibnamefont {Hattori}}, \bibinfo {author}
  {\bibfnamefont {N.}~\bibnamefont {Higa}}, \bibinfo {author} {\bibfnamefont
  {G.}~\bibnamefont {Nakamine}}, \emph {et~al.},\ }\bibfield  {title} {\bibinfo
  {title} {$^{125}${T}e-{N}{M}{R} {S}tudy on a {S}ingle {C}rystal of {H}eavy
  {F}ermion {S}uperconductor {U}{T}e$_2$},\ }\href@noop {} {\bibfield
  {journal} {\bibinfo  {journal} {J. Phys. Soc. Japan}\ }\textbf {\bibinfo
  {volume} {88}},\ \bibinfo {pages} {073701} (\bibinfo {year}
  {2019})}\BibitemShut {NoStop}%
\bibitem [{\citenamefont {Duan}\ \emph {et~al.}(2020)\citenamefont {Duan},
  \citenamefont {Sasmal}, \citenamefont {Maple}, \citenamefont {Podlesnyak},
  \citenamefont {Zhu}, \citenamefont {Si},\ and\ \citenamefont
  {Dai}}]{Duan_Arxiv_2020}%
  \BibitemOpen
  \bibfield  {author} {\bibinfo {author} {\bibfnamefont {C.}~\bibnamefont
  {Duan}}, \bibinfo {author} {\bibfnamefont {K.}~\bibnamefont {Sasmal}},
  \bibinfo {author} {\bibfnamefont {M.~B.}\ \bibnamefont {Maple}}, \bibinfo
  {author} {\bibfnamefont {A.}~\bibnamefont {Podlesnyak}}, \bibinfo {author}
  {\bibfnamefont {J.-X.}\ \bibnamefont {Zhu}}, \bibinfo {author} {\bibfnamefont
  {Q.}~\bibnamefont {Si}},\ and\ \bibinfo {author} {\bibfnamefont
  {P.}~\bibnamefont {Dai}},\ }\bibfield  {title} {\bibinfo {title}
  {Incommensurate {S}pin {F}luctuations in the {S}pin-{T}riplet
  {S}uperconductor {C}andidate {U}{T}e$_2$},\ }\href@noop {} {\bibfield
  {journal} {\bibinfo  {journal} {Phys. Rev. Lett.}\ }\textbf {\bibinfo
  {volume} {125}},\ \bibinfo {pages} {237003} (\bibinfo {year}
  {2020})}\BibitemShut {NoStop}%
\bibitem [{\citenamefont {Thomas}\ \emph {et~al.}(2020)\citenamefont {Thomas},
  \citenamefont {Santos}, \citenamefont {Christensen}, \citenamefont {Asaba},
  \citenamefont {Ronning}, \citenamefont {Thompson}, \citenamefont {Bauer},
  \citenamefont {Fernandes}, \citenamefont {Fabbris},\ and\ \citenamefont
  {Rosa}}]{Thomas_SciAdv_2020}%
  \BibitemOpen
  \bibfield  {author} {\bibinfo {author} {\bibfnamefont {S.~M.}\ \bibnamefont
  {Thomas}}, \bibinfo {author} {\bibfnamefont {F.~B.}\ \bibnamefont {Santos}},
  \bibinfo {author} {\bibfnamefont {M.~H.}\ \bibnamefont {Christensen}},
  \bibinfo {author} {\bibfnamefont {T.}~\bibnamefont {Asaba}}, \bibinfo
  {author} {\bibfnamefont {F.}~\bibnamefont {Ronning}}, \bibinfo {author}
  {\bibfnamefont {J.~D.}\ \bibnamefont {Thompson}}, \bibinfo {author}
  {\bibfnamefont {E.~D.}\ \bibnamefont {Bauer}}, \bibinfo {author}
  {\bibfnamefont {R.~M.}\ \bibnamefont {Fernandes}}, \bibinfo {author}
  {\bibfnamefont {G.}~\bibnamefont {Fabbris}},\ and\ \bibinfo {author}
  {\bibfnamefont {P.~F.~S.}\ \bibnamefont {Rosa}},\ }\bibfield  {title}
  {\bibinfo {title} {Evidence for a {P}ressure-{I}nduced {A}ntiferromagnetic
  {Q}uantum {C}ritical {P}oint in {I}ntermediate-{V}alence {U}{T}e$_{2}$},\
  }\href@noop {} {\bibfield  {journal} {\bibinfo  {journal} {Sci. Adv.}\
  }\textbf {\bibinfo {volume} {6}},\ \bibinfo {pages} {eabc8709} (\bibinfo
  {year} {2020})}\BibitemShut {NoStop}%
\bibitem [{\citenamefont {Aoki}\ \emph
  {et~al.}(2019{\natexlab{b}})\citenamefont {Aoki}, \citenamefont {Ishida},\
  and\ \citenamefont {Flouquet}}]{Aoki_JPSJ_2019}%
  \BibitemOpen
  \bibfield  {author} {\bibinfo {author} {\bibfnamefont {D.}~\bibnamefont
  {Aoki}}, \bibinfo {author} {\bibfnamefont {K.}~\bibnamefont {Ishida}},\ and\
  \bibinfo {author} {\bibfnamefont {J.}~\bibnamefont {Flouquet}},\ }\bibfield
  {title} {\bibinfo {title} {Review of {U}-{B}ased {F}erromagnetic
  {S}uperconductors: {C}omparison {B}etween {U}{G}e$_2$, {U}{R}h{G}e, and
  {U}{C}o{G}e},\ }\href@noop {} {\bibfield  {journal} {\bibinfo  {journal} {J.
  Phys. Soc. Japan}\ }\textbf {\bibinfo {volume} {88}},\ \bibinfo {pages}
  {022001} (\bibinfo {year} {2019}{\natexlab{b}})}\BibitemShut {NoStop}%
\bibitem [{\citenamefont {Pfleiderer}\ and\ \citenamefont
  {Huxley}(2002)}]{Pfleiderer_PRL_2002}%
  \BibitemOpen
  \bibfield  {author} {\bibinfo {author} {\bibfnamefont {C.}~\bibnamefont
  {Pfleiderer}}\ and\ \bibinfo {author} {\bibfnamefont {A.~D.}\ \bibnamefont
  {Huxley}},\ }\bibfield  {title} {\bibinfo {title} {Pressure {D}ependence of
  the {M}agnetization in the {F}erromagnetic {S}uperconductor {U}{G}e$_2$},\
  }\href@noop {} {\bibfield  {journal} {\bibinfo  {journal} {Phys. Rev. Lett.}\
  }\textbf {\bibinfo {volume} {89}},\ \bibinfo {pages} {147005} (\bibinfo
  {year} {2002})}\BibitemShut {NoStop}%
\bibitem [{\citenamefont {Levy}\ \emph {et~al.}(2005)\citenamefont {Levy},
  \citenamefont {Sheikin}, \citenamefont {Grenier},\ and\ \citenamefont
  {Huxley}}]{Levy_Science_2005}%
  \BibitemOpen
  \bibfield  {author} {\bibinfo {author} {\bibfnamefont {F.}~\bibnamefont
  {Levy}}, \bibinfo {author} {\bibfnamefont {I.}~\bibnamefont {Sheikin}},
  \bibinfo {author} {\bibfnamefont {B.}~\bibnamefont {Grenier}},\ and\ \bibinfo
  {author} {\bibfnamefont {A.~D.}\ \bibnamefont {Huxley}},\ }\bibfield  {title}
  {\bibinfo {title} {Magnetic {F}ield-{I}nduced {S}uperconductivity in the
  {F}erromagnet {U}{R}h{G}e},\ }\href@noop {} {\bibfield  {journal} {\bibinfo
  {journal} {Science}\ }\textbf {\bibinfo {volume} {309}},\ \bibinfo {pages}
  {1343} (\bibinfo {year} {2005})}\BibitemShut {NoStop}%
\bibitem [{\citenamefont {Ohta}\ \emph {et~al.}(2010)\citenamefont {Ohta},
  \citenamefont {Hattori}, \citenamefont {Ishida}, \citenamefont {Nakai},
  \citenamefont {Osaki}, \citenamefont {Deguchi}, \citenamefont {Sato},\ and\
  \citenamefont {Satoh}}]{Ohta_JPSJ_2010}%
  \BibitemOpen
  \bibfield  {author} {\bibinfo {author} {\bibfnamefont {T.}~\bibnamefont
  {Ohta}}, \bibinfo {author} {\bibfnamefont {T.}~\bibnamefont {Hattori}},
  \bibinfo {author} {\bibfnamefont {K.}~\bibnamefont {Ishida}}, \bibinfo
  {author} {\bibfnamefont {Y.}~\bibnamefont {Nakai}}, \bibinfo {author}
  {\bibfnamefont {E.}~\bibnamefont {Osaki}}, \bibinfo {author} {\bibfnamefont
  {K.}~\bibnamefont {Deguchi}}, \bibinfo {author} {\bibfnamefont {N.~K.}\
  \bibnamefont {Sato}},\ and\ \bibinfo {author} {\bibfnamefont
  {I.}~\bibnamefont {Satoh}},\ }\bibfield  {title} {\bibinfo {title}
  {Microscopic {C}oexistence of {F}erromagnetism and {S}uperconductivity in
  {S}ingle-{C}rystal {U}{C}o{G}e},\ }\href@noop {} {\bibfield  {journal}
  {\bibinfo  {journal} {J. Phys. Soc. Japan}\ }\textbf {\bibinfo {volume}
  {79}},\ \bibinfo {pages} {023707} (\bibinfo {year} {2010})}\BibitemShut
  {NoStop}%
\bibitem [{\citenamefont {Chevalier}\ \emph {et~al.}(1996)\citenamefont
  {Chevalier}, \citenamefont {Graverau}, \citenamefont {Berlureau},
  \citenamefont {Fournes},\ and\ \citenamefont
  {Etourneau}}]{Chevalier_AlloysAndCompounds_1996}%
  \BibitemOpen
  \bibfield  {author} {\bibinfo {author} {\bibfnamefont {B.}~\bibnamefont
  {Chevalier}}, \bibinfo {author} {\bibfnamefont {P.}~\bibnamefont {Graverau}},
  \bibinfo {author} {\bibfnamefont {T.}~\bibnamefont {Berlureau}}, \bibinfo
  {author} {\bibfnamefont {L.}~\bibnamefont {Fournes}},\ and\ \bibinfo {author}
  {\bibfnamefont {J.}~\bibnamefont {Etourneau}},\ }\bibfield  {title} {\bibinfo
  {title} {Ferromagnetic {P}roperties of {U}$_2${M}$_{17-y}${G}e$_y$},\
  }\href@noop {} {\bibfield  {journal} {\bibinfo  {journal} {J. Alloys
  Compounds}\ }\textbf {\bibinfo {volume} {233}},\ \bibinfo {pages} {174}
  (\bibinfo {year} {1996})}\BibitemShut {NoStop}%
\bibitem [{\citenamefont {Soude}\ \emph {et~al.}(2010)\citenamefont {Soude},
  \citenamefont {Tougait}, \citenamefont {Pasturel}, \citenamefont
  {Kaczorowski},\ and\ \citenamefont {Noel}}]{Soude_SolidStateChem_2010}%
  \BibitemOpen
  \bibfield  {author} {\bibinfo {author} {\bibfnamefont {A.}~\bibnamefont
  {Soude}}, \bibinfo {author} {\bibfnamefont {O.}~\bibnamefont {Tougait}},
  \bibinfo {author} {\bibfnamefont {M.}~\bibnamefont {Pasturel}}, \bibinfo
  {author} {\bibfnamefont {D.}~\bibnamefont {Kaczorowski}},\ and\ \bibinfo
  {author} {\bibfnamefont {H.}~\bibnamefont {Noel}},\ }\bibfield  {title}
  {\bibinfo {title} {Crystal {S}tructure and {E}lectronic {P}roperties of the
  {N}ew {C}ompounds {U}$_3${C}o$_{12-x}${X}$_4$ with {X}={S}i, {G}e},\
  }\href@noop {} {\bibfield  {journal} {\bibinfo  {journal} {J. Solid State
  Chem.}\ }\textbf {\bibinfo {volume} {183}},\ \bibinfo {pages} {1180}
  (\bibinfo {year} {2010})}\BibitemShut {NoStop}%
\bibitem [{\citenamefont {Bobev}\ \emph {et~al.}(2007)\citenamefont {Bobev},
  \citenamefont {Bauer}, \citenamefont {Ronning}, \citenamefont {Thompson},\
  and\ \citenamefont {Sarrao}}]{Bobev_SolidStateChem_2007}%
  \BibitemOpen
  \bibfield  {author} {\bibinfo {author} {\bibfnamefont {S.}~\bibnamefont
  {Bobev}}, \bibinfo {author} {\bibfnamefont {E.~D.}\ \bibnamefont {Bauer}},
  \bibinfo {author} {\bibfnamefont {F.}~\bibnamefont {Ronning}}, \bibinfo
  {author} {\bibfnamefont {J.~D.}\ \bibnamefont {Thompson}},\ and\ \bibinfo
  {author} {\bibfnamefont {J.~L.}\ \bibnamefont {Sarrao}},\ }\bibfield  {title}
  {\bibinfo {title} {Synthesis, {S}tructure and {P}hysical {P}roperties of the
  {N}ew {U}ranium {T}ernary {P}hase {U}$_3${C}o$_2${G}e$_7$},\ }\href@noop {}
  {\bibfield  {journal} {\bibinfo  {journal} {J. Solid State Chem.}\ }\textbf
  {\bibinfo {volume} {180}},\ \bibinfo {pages} {2830} (\bibinfo {year}
  {2007})}\BibitemShut {NoStop}%
\bibitem [{\citenamefont {Bauer}(2018)}]{Eric_Private_2018}%
  \BibitemOpen
  \bibfield  {author} {\bibinfo {author} {\bibfnamefont {E.~D.}\ \bibnamefont
  {Bauer}},\ }\href@noop {} {}\bibinfo {howpublished} {Personal Communication}
  (\bibinfo {year} {2018})\BibitemShut {NoStop}%
\bibitem [{\citenamefont {Wu}\ \emph {et~al.}(2018)\citenamefont {Wu},
  \citenamefont {Aoki},\ and\ \citenamefont {Brison}}]{Wu_PRB_2018}%
  \BibitemOpen
  \bibfield  {author} {\bibinfo {author} {\bibfnamefont {B.}~\bibnamefont
  {Wu}}, \bibinfo {author} {\bibfnamefont {D.}~\bibnamefont {Aoki}},\ and\
  \bibinfo {author} {\bibfnamefont {J.-P.}\ \bibnamefont {Brison}},\ }\bibfield
   {title} {\bibinfo {title} {Vortex {L}iquid {P}hase in the p-{W}ave
  {F}erromagnetic {S}uperconductor {U}{C}o{G}e},\ }\href@noop {} {\bibfield
  {journal} {\bibinfo  {journal} {Phys. Rev. B}\ }\textbf {\bibinfo {volume}
  {98}},\ \bibinfo {pages} {024517} (\bibinfo {year} {2018})}\BibitemShut
  {NoStop}%
\bibitem [{\citenamefont {Hattori}\ \emph {et~al.}(2012)\citenamefont
  {Hattori}, \citenamefont {Ihara}, \citenamefont {Nakai}, \citenamefont
  {Ishida}, \citenamefont {Tada}, \citenamefont {Fujimoto}, \citenamefont
  {Kawakami}, \citenamefont {Osaki}, \citenamefont {Deguchi}, \citenamefont
  {Sato},\ and\ \citenamefont {Satoh}}]{Hattori_PRL_2012}%
  \BibitemOpen
  \bibfield  {author} {\bibinfo {author} {\bibfnamefont {T.}~\bibnamefont
  {Hattori}}, \bibinfo {author} {\bibfnamefont {Y.}~\bibnamefont {Ihara}},
  \bibinfo {author} {\bibfnamefont {Y.}~\bibnamefont {Nakai}}, \bibinfo
  {author} {\bibfnamefont {K.}~\bibnamefont {Ishida}}, \bibinfo {author}
  {\bibfnamefont {Y.}~\bibnamefont {Tada}}, \bibinfo {author} {\bibfnamefont
  {S.}~\bibnamefont {Fujimoto}}, \bibinfo {author} {\bibfnamefont
  {N.}~\bibnamefont {Kawakami}}, \bibinfo {author} {\bibfnamefont
  {E.}~\bibnamefont {Osaki}}, \bibinfo {author} {\bibfnamefont
  {K.}~\bibnamefont {Deguchi}}, \bibinfo {author} {\bibfnamefont {N.~K.}\
  \bibnamefont {Sato}},\ and\ \bibinfo {author} {\bibfnamefont
  {I.}~\bibnamefont {Satoh}},\ }\bibfield  {title} {\bibinfo {title}
  {Superconductivity {I}nduced by {L}ongitudinal {F}erromagnetic {F}luctuations
  in {U}{C}o{G}e},\ }\href@noop {} {\bibfield  {journal} {\bibinfo  {journal}
  {Phys. Rev Lett.}\ }\textbf {\bibinfo {volume} {108}},\ \bibinfo {pages}
  {066403} (\bibinfo {year} {2012})}\BibitemShut {NoStop}%
\bibitem [{\citenamefont {Pospisil}\ \emph {et~al.}(2011)\citenamefont
  {Pospisil}, \citenamefont {Prokes}, \citenamefont {Reehuis}, \citenamefont
  {Tovar}, \citenamefont {Vejpravova}, \citenamefont {Prokleska},\ and\
  \citenamefont {Sechovsky}}]{Pospisil_JPSJ_2014}%
  \BibitemOpen
  \bibfield  {author} {\bibinfo {author} {\bibfnamefont {J.}~\bibnamefont
  {Pospisil}}, \bibinfo {author} {\bibfnamefont {K.}~\bibnamefont {Prokes}},
  \bibinfo {author} {\bibfnamefont {M.}~\bibnamefont {Reehuis}}, \bibinfo
  {author} {\bibfnamefont {M.}~\bibnamefont {Tovar}}, \bibinfo {author}
  {\bibfnamefont {J.~P.}\ \bibnamefont {Vejpravova}}, \bibinfo {author}
  {\bibfnamefont {J.}~\bibnamefont {Prokleska}},\ and\ \bibinfo {author}
  {\bibfnamefont {V.}~\bibnamefont {Sechovsky}},\ }\bibfield  {title} {\bibinfo
  {title} {Influence of {S}ample {P}reparation {T}echnology and {T}reatment on
  {M}agnetism and {S}uperconductivity of {U}{C}o{G}e},\ }\href@noop {}
  {\bibfield  {journal} {\bibinfo  {journal} {J. Phys. Soc. Japan}\ }\textbf
  {\bibinfo {volume} {80}},\ \bibinfo {pages} {084709} (\bibinfo {year}
  {2011})}\BibitemShut {NoStop}%
\bibitem [{\citenamefont {Kycia}(1997)}]{KyciaThesis}%
  \BibitemOpen
  \bibfield  {author} {\bibinfo {author} {\bibfnamefont {J.~B.}\ \bibnamefont
  {Kycia}},\ }\emph {\bibinfo {title} {Growth and {C}haracterization of {H}igh
  {Q}uality {U}{P}t$_3$ {S}ingle {C}rystals and {H}igh {R}esolution {N}{M}{R}
  {S}tudy of {S}uperfluid $^3${H}e-{B}}},\ \href@noop {} {Ph.D. thesis},\
  \bibinfo  {school} {Northwestern University}, \bibinfo {address} {2145
  Sheridan Road Evanston, IL 60208} (\bibinfo {year} {1997})\BibitemShut
  {NoStop}%
\bibitem [{\citenamefont {Kycia}\ \emph {et~al.}(1998)\citenamefont {Kycia},
  \citenamefont {Hong}, \citenamefont {Graf}, \citenamefont {Sauls},
  \citenamefont {Seidman},\ and\ \citenamefont {Halperin}}]{Kycia_PRB_1998}%
  \BibitemOpen
  \bibfield  {author} {\bibinfo {author} {\bibfnamefont {J.~B.}\ \bibnamefont
  {Kycia}}, \bibinfo {author} {\bibfnamefont {J.~I.}\ \bibnamefont {Hong}},
  \bibinfo {author} {\bibfnamefont {M.~J.}\ \bibnamefont {Graf}}, \bibinfo
  {author} {\bibfnamefont {J.~A.}\ \bibnamefont {Sauls}}, \bibinfo {author}
  {\bibfnamefont {D.~N.}\ \bibnamefont {Seidman}},\ and\ \bibinfo {author}
  {\bibfnamefont {W.~P.}\ \bibnamefont {Halperin}},\ }\bibfield  {title}
  {\bibinfo {title} {Suppression of {S}uperconductivity in {U}{P}t$_3$ {S}ingle
  {C}rystals},\ }\href@noop {} {\bibfield  {journal} {\bibinfo  {journal}
  {Phys. Rev. B}\ }\textbf {\bibinfo {volume} {58}},\ \bibinfo {pages} {R603}
  (\bibinfo {year} {1998})}\BibitemShut {NoStop}%
\bibitem [{\citenamefont {Joynt}\ and\ \citenamefont
  {Taillefer}(2002)}]{Joynt_RevModPhys_2002}%
  \BibitemOpen
  \bibfield  {author} {\bibinfo {author} {\bibfnamefont {R.}~\bibnamefont
  {Joynt}}\ and\ \bibinfo {author} {\bibfnamefont {L.}~\bibnamefont
  {Taillefer}},\ }\bibfield  {title} {\bibinfo {title} {The {S}uperconducting
  {P}hases of {U}{P}t$_3$},\ }\href@noop {} {\bibfield  {journal} {\bibinfo
  {journal} {Reviews of Modern Physics}\ }\textbf {\bibinfo {volume} {74}},\
  \bibinfo {pages} {235} (\bibinfo {year} {2002})}\BibitemShut {NoStop}%
\bibitem [{\citenamefont {Gasparini}\ \emph {et~al.}(2010)\citenamefont
  {Gasparini}, \citenamefont {Huang}, \citenamefont {Huy}, \citenamefont
  {Klaasse}, \citenamefont {Naka}, \citenamefont {Slooten},\ and\ \citenamefont
  {deVisser}}]{Gasparini_LowTempPhys_2010}%
  \BibitemOpen
  \bibfield  {author} {\bibinfo {author} {\bibfnamefont {A.}~\bibnamefont
  {Gasparini}}, \bibinfo {author} {\bibfnamefont {Y.~K.}\ \bibnamefont
  {Huang}}, \bibinfo {author} {\bibfnamefont {N.~T.}\ \bibnamefont {Huy}},
  \bibinfo {author} {\bibfnamefont {J.~C.~P.}\ \bibnamefont {Klaasse}},
  \bibinfo {author} {\bibfnamefont {T.}~\bibnamefont {Naka}}, \bibinfo {author}
  {\bibfnamefont {E.}~\bibnamefont {Slooten}},\ and\ \bibinfo {author}
  {\bibfnamefont {A.}~\bibnamefont {deVisser}},\ }\bibfield  {title} {\bibinfo
  {title} {The {S}uperconductor {F}erromagnet {U}{C}o{G}e},\ }\href@noop {}
  {\bibfield  {journal} {\bibinfo  {journal} {J. of Low Temp. Phys.}\ }\textbf
  {\bibinfo {volume} {161}},\ \bibinfo {pages} {134} (\bibinfo {year}
  {2010})}\BibitemShut {NoStop}%
\bibitem [{\citenamefont {Taupin}\ \emph {et~al.}(2014)\citenamefont {Taupin},
  \citenamefont {Howald}, \citenamefont {Aoki},\ and\ \citenamefont
  {Brison}}]{Taupin_PRB_2014}%
  \BibitemOpen
  \bibfield  {author} {\bibinfo {author} {\bibfnamefont {M.}~\bibnamefont
  {Taupin}}, \bibinfo {author} {\bibfnamefont {L.}~\bibnamefont {Howald}},
  \bibinfo {author} {\bibfnamefont {D.}~\bibnamefont {Aoki}},\ and\ \bibinfo
  {author} {\bibfnamefont {J.-P.}\ \bibnamefont {Brison}},\ }\bibfield  {title}
  {\bibinfo {title} {Superconducting {G}ap of {U}{C}o{G}e {P}robed by {T}hermal
  {T}ransport},\ }\href@noop {} {\bibfield  {journal} {\bibinfo  {journal}
  {Phys. Rev. B}\ }\textbf {\bibinfo {volume} {90}},\ \bibinfo {pages}
  {180501(R)} (\bibinfo {year} {2014})}\BibitemShut {NoStop}%
\bibitem [{\citenamefont {Hewson}(1993)}]{hewson_1993}%
  \BibitemOpen
  \bibfield  {author} {\bibinfo {author} {\bibfnamefont {A.~C.}\ \bibnamefont
  {Hewson}},\ }\href {https://doi.org/10.1017/CBO9780511470752} {\emph
  {\bibinfo {title} {The Kondo Problem to Heavy Fermions}}},\ Cambridge Studies
  in Magnetism\ (\bibinfo  {publisher} {Cambridge University Press},\ \bibinfo
  {year} {1993})\BibitemShut {NoStop}%
\bibitem [{\citenamefont {Brando}\ \emph {et~al.}(2016)\citenamefont {Brando},
  \citenamefont {Belitz}, \citenamefont {Grosche},\ and\ \citenamefont
  {Kirkpatrick}}]{Brando_RevModPhys_2002}%
  \BibitemOpen
  \bibfield  {author} {\bibinfo {author} {\bibfnamefont {M.}~\bibnamefont
  {Brando}}, \bibinfo {author} {\bibfnamefont {D.}~\bibnamefont {Belitz}},
  \bibinfo {author} {\bibfnamefont {F.~M.}\ \bibnamefont {Grosche}},\ and\
  \bibinfo {author} {\bibfnamefont {T.~R.}\ \bibnamefont {Kirkpatrick}},\
  }\bibfield  {title} {\bibinfo {title} {Metallic {Q}uantum {F}erromagnets},\
  }\href@noop {} {\bibfield  {journal} {\bibinfo  {journal} {Rev. Mod. Phys.}\
  }\textbf {\bibinfo {volume} {88}},\ \bibinfo {pages} {025006} (\bibinfo
  {year} {2016})}\BibitemShut {NoStop}%
\bibitem [{\citenamefont {Huy}\ \emph {et~al.}(2009)\citenamefont {Huy},
  \citenamefont {Huang},\ and\ \citenamefont
  {deVisser}}]{Huy_MagMaterials_2009}%
  \BibitemOpen
  \bibfield  {author} {\bibinfo {author} {\bibfnamefont {N.~T.}\ \bibnamefont
  {Huy}}, \bibinfo {author} {\bibfnamefont {Y.~K.}\ \bibnamefont {Huang}},\
  and\ \bibinfo {author} {\bibfnamefont {A.}~\bibnamefont {deVisser}},\
  }\bibfield  {title} {\bibinfo {title} {Effect of {A}nnealing on the
  {M}agnetic and {S}uperconducting {P}roperties of {S}ingle-{C}rystalline
  {U}{C}o{G}e},\ }\href@noop {} {\bibfield  {journal} {\bibinfo  {journal} {J.
  Magn. Magn. Mater.}\ }\textbf {\bibinfo {volume} {321}},\ \bibinfo {pages}
  {2691} (\bibinfo {year} {2009})}\BibitemShut {NoStop}%
\bibitem [{\citenamefont {Jacko}\ \emph {et~al.}(2009)\citenamefont {Jacko},
  \citenamefont {Fjaerestad},\ and\ \citenamefont
  {Powell}}]{Jacko_Nature_2009}%
  \BibitemOpen
  \bibfield  {author} {\bibinfo {author} {\bibfnamefont {A.~C.}\ \bibnamefont
  {Jacko}}, \bibinfo {author} {\bibfnamefont {J.~O.}\ \bibnamefont
  {Fjaerestad}},\ and\ \bibinfo {author} {\bibfnamefont {B.~J.}\ \bibnamefont
  {Powell}},\ }\bibfield  {title} {\bibinfo {title} {A {U}nified {E}xplanation
  of the {K}adowaki-{W}oods {R}atio in the {S}trongly {C}orrelated {M}etals.},\
  }\href@noop {} {\bibfield  {journal} {\bibinfo  {journal} {Nature Phys.}\
  }\textbf {\bibinfo {volume} {5}},\ \bibinfo {pages} {422} (\bibinfo {year}
  {2009})}\BibitemShut {NoStop}%
\bibitem [{\citenamefont {Aoki}\ and\ \citenamefont
  {Flouquet}(2014)}]{Aoki_JPSJ_2014}%
  \BibitemOpen
  \bibfield  {author} {\bibinfo {author} {\bibfnamefont {D.}~\bibnamefont
  {Aoki}}\ and\ \bibinfo {author} {\bibfnamefont {J.}~\bibnamefont
  {Flouquet}},\ }\bibfield  {title} {\bibinfo {title} {Superconductivity and
  {F}erromagentic {Q}uantum {C}riticality in {U}ranium {C}ompounds},\
  }\href@noop {} {\bibfield  {journal} {\bibinfo  {journal} {J. Phys. Soc.
  Japan}\ }\textbf {\bibinfo {volume} {83}},\ \bibinfo {pages} {061011}
  (\bibinfo {year} {2014})}\BibitemShut {NoStop}%
\bibitem [{\citenamefont {Hardy}\ \emph {et~al.}(2011)\citenamefont {Hardy},
  \citenamefont {Aoki}, \citenamefont {Meingast}, \citenamefont {Schweiss},
  \citenamefont {Burger}, \citenamefont {v.~Lohneysen},\ and\ \citenamefont
  {Flouquet}}]{Hardy_PRB_2011}%
  \BibitemOpen
  \bibfield  {author} {\bibinfo {author} {\bibfnamefont {F.}~\bibnamefont
  {Hardy}}, \bibinfo {author} {\bibfnamefont {D.}~\bibnamefont {Aoki}},
  \bibinfo {author} {\bibfnamefont {C.}~\bibnamefont {Meingast}}, \bibinfo
  {author} {\bibfnamefont {P.}~\bibnamefont {Schweiss}}, \bibinfo {author}
  {\bibfnamefont {P.}~\bibnamefont {Burger}}, \bibinfo {author} {\bibfnamefont
  {H.}~\bibnamefont {v.~Lohneysen}},\ and\ \bibinfo {author} {\bibfnamefont
  {J.}~\bibnamefont {Flouquet}},\ }\bibfield  {title} {\bibinfo {title}
  {Transverse and {L}ongitudinal {M}agnetic-field {R}esponse in the {I}sing
  {F}erromagnets {U}{R}h{G}e, {U}{C}o{G}e, and {U}{G}e$_2$},\ }\href@noop {}
  {\bibfield  {journal} {\bibinfo  {journal} {Phys. Rev. B}\ }\textbf {\bibinfo
  {volume} {83}},\ \bibinfo {pages} {195107} (\bibinfo {year}
  {2011})}\BibitemShut {NoStop}%
\bibitem [{\citenamefont {Pospisil}\ \emph {et~al.}(2020)\citenamefont
  {Pospisil}, \citenamefont {Haga}, \citenamefont {Miyake}, \citenamefont
  {Kambe}, \citenamefont {Tokunaga}, \citenamefont {Tokunaga}, \citenamefont
  {Yamamoto}, \citenamefont {Proschek}, \citenamefont {Volny},\ and\
  \citenamefont {Sechovsky}}]{Pospisil_PRB_2020}%
  \BibitemOpen
  \bibfield  {author} {\bibinfo {author} {\bibfnamefont {J.}~\bibnamefont
  {Pospisil}}, \bibinfo {author} {\bibfnamefont {Y.}~\bibnamefont {Haga}},
  \bibinfo {author} {\bibfnamefont {A.}~\bibnamefont {Miyake}}, \bibinfo
  {author} {\bibfnamefont {S.}~\bibnamefont {Kambe}}, \bibinfo {author}
  {\bibfnamefont {Y.}~\bibnamefont {Tokunaga}}, \bibinfo {author}
  {\bibfnamefont {M.}~\bibnamefont {Tokunaga}}, \bibinfo {author}
  {\bibfnamefont {E.}~\bibnamefont {Yamamoto}}, \bibinfo {author}
  {\bibfnamefont {P.}~\bibnamefont {Proschek}}, \bibinfo {author}
  {\bibfnamefont {J.}~\bibnamefont {Volny}},\ and\ \bibinfo {author}
  {\bibfnamefont {V.}~\bibnamefont {Sechovsky}},\ }\bibfield  {title} {\bibinfo
  {title} {Intriguing {B}ehavior of {U}{C}o$_{1-x}${R}h$_x${G}e {F}erromagnets
  in {M}agnetic {F}ield {A}long the b {A}xis},\ }\href@noop {} {\bibfield
  {journal} {\bibinfo  {journal} {Phys. Rev. B}\ }\textbf {\bibinfo {volume}
  {102}},\ \bibinfo {pages} {024442} (\bibinfo {year} {2020})}\BibitemShut
  {NoStop}%
\bibitem [{\citenamefont {Aoki}\ and\ \citenamefont
  {Flouquet}(2012)}]{Aoki_JPSJ_2012}%
  \BibitemOpen
  \bibfield  {author} {\bibinfo {author} {\bibfnamefont {D.}~\bibnamefont
  {Aoki}}\ and\ \bibinfo {author} {\bibfnamefont {J.}~\bibnamefont
  {Flouquet}},\ }\bibfield  {title} {\bibinfo {title} {Ferromagnetism and
  {S}uperconductivity in {U}ranium {C}ompounds},\ }\href@noop {} {\bibfield
  {journal} {\bibinfo  {journal} {J. Phys. Soc. Japan}\ }\textbf {\bibinfo
  {volume} {81}},\ \bibinfo {pages} {011003} (\bibinfo {year}
  {2012})}\BibitemShut {NoStop}%
\bibitem [{\citenamefont {Aoki}\ \emph {et~al.}(2011)\citenamefont {Aoki},
  \citenamefont {Sheikin}, \citenamefont {Matsuda}, \citenamefont {Taufour},
  \citenamefont {Knebel},\ and\ \citenamefont {Flouquet}}]{Aoki_JPSJ_2011}%
  \BibitemOpen
  \bibfield  {author} {\bibinfo {author} {\bibfnamefont {D.}~\bibnamefont
  {Aoki}}, \bibinfo {author} {\bibfnamefont {I.}~\bibnamefont {Sheikin}},
  \bibinfo {author} {\bibfnamefont {T.~D.}\ \bibnamefont {Matsuda}}, \bibinfo
  {author} {\bibfnamefont {V.}~\bibnamefont {Taufour}}, \bibinfo {author}
  {\bibfnamefont {G.}~\bibnamefont {Knebel}},\ and\ \bibinfo {author}
  {\bibfnamefont {J.}~\bibnamefont {Flouquet}},\ }\bibfield  {title} {\bibinfo
  {title} {First {O}bservation of {Q}uantum {O}scillations in the
  {F}erromagnetic {S}uperconductor {U}{C}o{G}e},\ }\href@noop {} {\bibfield
  {journal} {\bibinfo  {journal} {J. Phys. Soc. Japan}\ }\textbf {\bibinfo
  {volume} {80}},\ \bibinfo {pages} {013705} (\bibinfo {year}
  {2011})}\BibitemShut {NoStop}%
\bibitem [{\citenamefont {Bay}\ \emph {et~al.}(2014)\citenamefont {Bay},
  \citenamefont {Nikitin}, \citenamefont {Naka}, \citenamefont {McCollam},
  \citenamefont {Huang},\ and\ \citenamefont {deVisser}}]{Bay_PRB_2014}%
  \BibitemOpen
  \bibfield  {author} {\bibinfo {author} {\bibfnamefont {T.~V.}\ \bibnamefont
  {Bay}}, \bibinfo {author} {\bibfnamefont {A.~M.}\ \bibnamefont {Nikitin}},
  \bibinfo {author} {\bibfnamefont {T.}~\bibnamefont {Naka}}, \bibinfo {author}
  {\bibfnamefont {A.}~\bibnamefont {McCollam}}, \bibinfo {author}
  {\bibfnamefont {Y.~K.}\ \bibnamefont {Huang}},\ and\ \bibinfo {author}
  {\bibfnamefont {A.}~\bibnamefont {deVisser}},\ }\bibfield  {title} {\bibinfo
  {title} {Angular {V}ariation of the {M}agnetoresistance of the
  {S}uperconducting {F}erromagnet {U}{C}o{G}e},\ }\href@noop {} {\bibfield
  {journal} {\bibinfo  {journal} {Phys. Rev. B}\ }\textbf {\bibinfo {volume}
  {89}},\ \bibinfo {pages} {214512} (\bibinfo {year} {2014})}\BibitemShut
  {NoStop}%
\bibitem [{\citenamefont {Bastien}\ \emph {et~al.}(2016)\citenamefont
  {Bastien}, \citenamefont {Gourgort}, \citenamefont {Aoki}, \citenamefont
  {Pourret}, \citenamefont {Sheikin}, \citenamefont {Seyfarth}, \citenamefont
  {Flouquet},\ and\ \citenamefont {Knebel}}]{Bastien_PRL_2016}%
  \BibitemOpen
  \bibfield  {author} {\bibinfo {author} {\bibfnamefont {G.}~\bibnamefont
  {Bastien}}, \bibinfo {author} {\bibfnamefont {A.}~\bibnamefont {Gourgort}},
  \bibinfo {author} {\bibfnamefont {D.}~\bibnamefont {Aoki}}, \bibinfo {author}
  {\bibfnamefont {A.}~\bibnamefont {Pourret}}, \bibinfo {author} {\bibfnamefont
  {I.}~\bibnamefont {Sheikin}}, \bibinfo {author} {\bibfnamefont
  {G.}~\bibnamefont {Seyfarth}}, \bibinfo {author} {\bibfnamefont
  {J.}~\bibnamefont {Flouquet}},\ and\ \bibinfo {author} {\bibfnamefont
  {G.}~\bibnamefont {Knebel}},\ }\bibfield  {title} {\bibinfo {title} {Lifshitz
  {T}ransitions in the {F}erromagnetic {S}uperconductor {U}{C}o{G}e},\
  }\href@noop {} {\bibfield  {journal} {\bibinfo  {journal} {Phys. Rev. Lett.}\
  }\textbf {\bibinfo {volume} {117}},\ \bibinfo {pages} {206401} (\bibinfo
  {year} {2016})}\BibitemShut {NoStop}%
\bibitem [{\citenamefont {Knafo}\ \emph {et~al.}(2012)\citenamefont {Knafo},
  \citenamefont {Matsuda}, \citenamefont {Aoki}, \citenamefont {Hardy},
  \citenamefont {Scheerer}, \citenamefont {Ballon}, \citenamefont {Nardone},
  \citenamefont {Zitouni}, \citenamefont {Meingast},\ and\ \citenamefont
  {Flouquet}}]{Knafo_PRB_2012}%
  \BibitemOpen
  \bibfield  {author} {\bibinfo {author} {\bibfnamefont {W.}~\bibnamefont
  {Knafo}}, \bibinfo {author} {\bibfnamefont {T.~D.}\ \bibnamefont {Matsuda}},
  \bibinfo {author} {\bibfnamefont {D.}~\bibnamefont {Aoki}}, \bibinfo {author}
  {\bibfnamefont {F.}~\bibnamefont {Hardy}}, \bibinfo {author} {\bibfnamefont
  {G.~W.}\ \bibnamefont {Scheerer}}, \bibinfo {author} {\bibfnamefont
  {G.}~\bibnamefont {Ballon}}, \bibinfo {author} {\bibfnamefont
  {M.}~\bibnamefont {Nardone}}, \bibinfo {author} {\bibfnamefont
  {A.}~\bibnamefont {Zitouni}}, \bibinfo {author} {\bibfnamefont
  {C.}~\bibnamefont {Meingast}},\ and\ \bibinfo {author} {\bibfnamefont
  {J.}~\bibnamefont {Flouquet}},\ }\bibfield  {title} {\bibinfo {title}
  {High-field {M}oment {P}olarization in the {F}erromagnetic {S}uperconductor
  {U}{C}o{G}e},\ }\href@noop {} {\bibfield  {journal} {\bibinfo  {journal}
  {Phys. Rev. B}\ }\textbf {\bibinfo {volume} {86}},\ \bibinfo {pages} {184416}
  (\bibinfo {year} {2012})}\BibitemShut {NoStop}%
\bibitem [{\citenamefont {Gorkov}(1987)}]{Gorkov_SovSciRevA_1987}%
  \BibitemOpen
  \bibfield  {author} {\bibinfo {author} {\bibfnamefont {L.~P.}\ \bibnamefont
  {Gorkov}},\ }\bibfield  {title} {\bibinfo {title} {Superconductivity in
  {H}eavy {F}ermion {S}ystems},\ }\href@noop {} {\bibfield  {journal} {\bibinfo
   {journal} {Sov. Sci. Rev. A}\ }\textbf {\bibinfo {volume} {9}},\ \bibinfo
  {pages} {1} (\bibinfo {year} {1987})}\BibitemShut {NoStop}%
\bibitem [{\citenamefont {Troc}\ \emph {et~al.}(2010)\citenamefont {Troc},
  \citenamefont {Wawryk}, \citenamefont {Miiller}, \citenamefont {Misiorek},\
  and\ \citenamefont {Samsel-Czekala}}]{Troc_PM_2010}%
  \BibitemOpen
  \bibfield  {author} {\bibinfo {author} {\bibfnamefont {R.}~\bibnamefont
  {Troc}}, \bibinfo {author} {\bibfnamefont {R.}~\bibnamefont {Wawryk}},
  \bibinfo {author} {\bibfnamefont {W.}~\bibnamefont {Miiller}}, \bibinfo
  {author} {\bibfnamefont {H.}~\bibnamefont {Misiorek}},\ and\ \bibinfo
  {author} {\bibfnamefont {M.}~\bibnamefont {Samsel-Czekala}},\ }\bibfield
  {title} {\bibinfo {title} {Bulk {P}roperties of the {U}{C}o{G}e
  {K}ondo-{L}ike {S}ystem},\ }\href@noop {} {\bibfield  {journal} {\bibinfo
  {journal} {Philos. Mag.}\ }\textbf {\bibinfo {volume} {90}},\ \bibinfo
  {pages} {2249} (\bibinfo {year} {2010})}\BibitemShut {NoStop}%
\bibitem [{\citenamefont {Huy}\ \emph {et~al.}(2007)\citenamefont {Huy},
  \citenamefont {Gasparini}, \citenamefont {de~Nijs}, \citenamefont {Huang},
  \citenamefont {Klaasse}, \citenamefont {Gortenmulder}, \citenamefont
  {deVisser}, \citenamefont {Hamann}, \citenamefont {Gorlach},\ and\
  \citenamefont {v.~Lohneysen}}]{Huy_PRL_2007}%
  \BibitemOpen
  \bibfield  {author} {\bibinfo {author} {\bibfnamefont {N.~T.}\ \bibnamefont
  {Huy}}, \bibinfo {author} {\bibfnamefont {A.}~\bibnamefont {Gasparini}},
  \bibinfo {author} {\bibfnamefont {D.~E.}\ \bibnamefont {de~Nijs}}, \bibinfo
  {author} {\bibfnamefont {Y.}~\bibnamefont {Huang}}, \bibinfo {author}
  {\bibfnamefont {J.~C.~P.}\ \bibnamefont {Klaasse}}, \bibinfo {author}
  {\bibfnamefont {T.}~\bibnamefont {Gortenmulder}}, \bibinfo {author}
  {\bibfnamefont {A.}~\bibnamefont {deVisser}}, \bibinfo {author}
  {\bibfnamefont {A.}~\bibnamefont {Hamann}}, \bibinfo {author} {\bibfnamefont
  {T.}~\bibnamefont {Gorlach}},\ and\ \bibinfo {author} {\bibfnamefont
  {H.}~\bibnamefont {v.~Lohneysen}},\ }\bibfield  {title} {\bibinfo {title}
  {Superconductivity on the {B}order of {W}eak {I}tinerant {F}erromagnetism in
  {U}{C}o{G}e},\ }\href@noop {} {\bibfield  {journal} {\bibinfo  {journal}
  {Phys. Rev. Lett.}\ }\textbf {\bibinfo {volume} {99}},\ \bibinfo {pages}
  {067006} (\bibinfo {year} {2007})}\BibitemShut {NoStop}%
\end{thebibliography}
%
\end{document}